\documentclass{article}
\PassOptionsToPackage{numbers, compress}{natbib}
\usepackage[preprint]{neurips_2021}
\usepackage[utf8]{inputenc} 
\usepackage[T1]{fontenc}    
\usepackage{hyperref}       
\hypersetup{
    colorlinks=true,
    linkcolor=blue,
    filecolor=magenta,      
    urlcolor=cyan
    }
\usepackage{url}            
\usepackage{booktabs}       
\usepackage{amsfonts}       
\usepackage{nicefrac}       
\usepackage{microtype}      
\usepackage{xcolor}         
\usepackage{graphicx}
\usepackage{mathtools}
\usepackage{amssymb}
\usepackage{amsmath}
\usepackage[ruled,vlined]{algorithm2e}
\usepackage{multicol}
\usepackage{multirow}
\usepackage{subfig}

\usepackage{amsmath,hyperref}
\usepackage{graphicx}
\usepackage{breqn}
\usepackage[ruled,vlined]{algorithm2e}
\usepackage{amsfonts}
\usepackage{amsthm}

\newtheorem{hyp}{Hypothesis}

\DeclareMathOperator*{\argmin}{argmin}
\usepackage[export]{adjustbox}          
\usepackage{cleveref}
\usepackage{xcolor}
\usepackage{wrapfig}

\usepackage{graphicx}
\usepackage[disable]{todonotes}
\usepackage{caption}
\usepackage{balance}
\usepackage{wrapfig}
\usepackage{float}
\title{Epidemic Control Modeling using Parsimonious Models and Markov Decision Processes}

\author{%
  Edilson~F. Arruda\\
  University of Southampton, UK\\
   \texttt{E.F.Arruda@southampton.ac.uk}\\
  \And
  Tarun~Sharma\\
  Number Theory Software Pvt Ltd, Delhi\\
  \texttt{tarun81998@gmail.com}\\
     \And
  Rodrigo~e~A.~Alexandre\\
  Federal University of Rio de Janeiro, Brasil\\
    \texttt{alvim.rodrigo@yahoo.com.br}
  \And
  Sinnu~Susan~Thomas\\
  Digital University Kerala, India\\
  \texttt{sinnu.thomas@duk.ac.in} \\
}

\begin{document}

\maketitle

\begin{abstract}

Many countries have experienced at least two waves of the COVID-19 pandemic. The second wave is far more dangerous as distinct strains appear more harmful to human health, but it stems from the complacency about the first wave. This paper introduces a parsimonious yet representative stochastic epidemic model that simulates the uncertain spread of the disease regardless of the latency and recovery time distributions. We also propose a  Markov decision process to seek an optimal trade-off between the usage of the healthcare system and the economic costs of an epidemic. We apply the model to COVID-19 data from New Delhi, India and simulate the epidemic spread with different policy review times. The results show that the optimal policy acts swiftly to curb the epidemic in the first wave, thus avoiding the collapse of the healthcare system and the future costs of posterior outbreaks. An analysis of the recent collapse of the healthcare system of India during the second COVID-19 wave suggests that many lives could have been preserved if swift mitigation was promoted after the first wave.
\end{abstract}

\section{Introduction}\label{sec:introduction}
Since the beginning of 2020, the COVID-19 pandemic has been challenging decision-making around the globe. Reinfection, multiple viral strains and inconsistent reporting are just some of the issues that decision-makers ought to consider \cite{Arruda2021-Plos,Dyern2021,Sabino2021}. Another important component is the geographical scope of the mitigating policies. Although the pandemic is a globally interconnected event and despite calls for unified response efforts \cite{Priesemann2021,Priesemann2021a}, mitigating measures are generally planned at a national or local level.

To properly prescribe containment measures, decision makers must consider not only the current spread, but also the future consequences of such measures. These consequences include immediate economic losses, as well as the costs and effects of further measures that might result from ineffective containment. An ineffective mitigation may lead to high prevalence which, in turn, requires stronger and therefore more costly mitigation \cite{Arruda2021-Plos,Priesemann2021a}. Furthermore, under high prevalence, mitigation measures become increasingly expensive and difficult to manage and enforce \cite{Priesemann2021a}. Hence, it is clear that mitigating policies should be designed to manage both the infection levels, as well as the mitigation costs and future consequences. Since all of these components involve a large amount of uncertainty, this paper will use a stochastic approach to model the spread and the effects of mitigation.

Vital to the modelling of outbreaks, classical epidemic models \cite{Ross1916,Kermack1927} use deterministic differential equations to describe the spread and are sensitive  to uncertain disease-specific parameters \cite{Walling2007}.  Stochastic models were introduced later \cite{Allen2008} to consider the underlying uncertainties in the spread and in the duration of the disease cycle \cite{Backer2020}. However, analytical and computational tractability is vital to append optimisation under uncertainty into the framework and provide decision support for mitigation.

Making use of queuing systems with infinitely many servers, we introduce a parsimonious yet realistic stochastic epidemic model that considers general latency and infection times. We then introduce a Markov decision process to derive optimal control actions \cite{RePEc:eee:chsofr:v:136:y:2020:i:c:s0960077920302897} that seek a balance between the occupation of the healthcare system and economic costs, depending on the current number of infected and exposed individuals. We apply the approach to COVID-19 data from New Delhi and investigate the controlled output of the epidemic model. The results illustrate the importance of acting swiftly to contain the epidemic and avoid multiple waves, as well as the importance of continuously monitoring the outbreak as larger control review intervals imply higher optimal control levels.

The remainder of this paper is organised as follows. Section \ref{sec:review} contains a brief literature review and underscores the proposed contributions. Section \ref{sec:model} introduces the model and the optimal control formulation. Section \ref{sec:experiments} features experimental discussion based on the COVID-19 epidemic in New Delhi. Finally, Section \ref{sec:conc} concludes the paper.

\section{Literature Review \label{sec:review}}

Early in the COVID-19 pandemic, an influential study helped shape public policy by championing \emph{non-pharmaceutical interventions} (NPI) such as social isolation, home quarantine and lock-downs \cite{ferguson2020}. Intended to preserve the healthcare system, these measures also have a profound economic impact \cite{You2020}, which may lead policy makers to costly ill-timed interventions \cite{Oraby2021,Tarrataca2021}. 

Understanding the spread is vital for decision support. To that end, authors employed artificial intelligence to assess the epidemic threshold, predict the spread and reconcile classical epidemic models with the wealth of epidemic data made available. The same approach was applied to evaluate specific mitigation measures in distinct localities \cite{cheetham2020determining,bergonzi2020discrete}. While useful, data-based models are hindered by the quality of the reported data \cite{Latif2020}, which is often inconsistent as it depends on testing and reporting policies that differ widely around the globe \cite{Tarrataca2021,Dyern2021}; furthermore, even a quality testing strategy can be compromised by the uncertain and reportedly large percentage of asymptomatic infections generally not captured in the reports \cite{Meyerowitz2020}.

The trade-off between societal and economic consequences motivated discussions on the ideal duration and intensity of lock-downs \cite{plazas2021modeling,Tarrataca2021}. An analytic approach is to use optimal control within a classical epidemic model to seek a balance between societal and economic costs \cite{Kantner2020,Perkins2020,Arruda2021-Plos}. While useful, these models provide only a predefined open-loop strategy that disregards the stochastic variations of the spread.

Time series predictive models can incorporate observed fluctuations in the data \cite{masum2020r}, but do not benefit from analytic epidemic models. To emulate uncertain propagation within a network, a Monte Carlo Markov chain approach considers social media interactions. Similar network models can evaluate on-off quarantine strategies \cite{meidan2021alternating}, or predict the geographical spread of a disease with the dynamics underpinned by a classical epidemic model. Perhaps because of their complexity, these models do not embed optimization in the design of intervention strategies. To include optimization, \cite{ohi2020exploring} integrated reinforcement learning and agent-based simulation within a virtual environment. The model is computationally expensive and therefore limited to small population sizes.

Parsimonious models such as SEIR (Susceptible, Exposed, Infected, Removed) effectively described the 1918 flu epidemic  and the COVID-19 epidemic in the US \cite{Bootsma2007,Bertozzi2020}. Computationally effective and analytically tractable, these models provide an ideal framework for optimal control \cite{Kantner2020,Perkins2020}. To derive a closed-loop strategy whilst considering an uncertain disease spread, we utilize the stochastic SEIR framework \cite{Allen2008,Britton2010}. Markov processes underpin stochastic SEIR models \cite{Artalejo2015,Lopez2017,Amador2018} that assume exponentially distributed latency and infection periods. \cite{Artalejo2015} studies the outbreak's duration, \cite{Lopez2017} derives the transmission rate per infected person and \cite{Amador2018} estimates the epidemic's size. General infection periods demand less tractable semi-Markov models \cite{Clancy2014,Corral2017}. These are intended to predict the outbreak and do not include optimization.

The contributions of this paper are twofold. Firstly, we introduce a parsimonious stochastic model of epidemic progression that seamlessly incorporates general latency and infectious periods. In contrast to the complex and computationally expensive neural network designs in \cite{ohi2020exploring}, we use a simple $M_t/G/\infty$ whose output at any time is a Poisson variable \cite{Eick1993}. Although as general as the semi-Markov models in \cite{Clancy2014,Corral2017}, the model remains tractable for large population sizes. Secondly, we use Markov decision processes \cite{Puterman1994} to introduce a stochastic optimal control formulation that seeks an optimal trade-off between the healthcare system's occupation and the economic impacts of mitigation measures. Since the model is based on parsimonious stochastic formulations, it is tractable, easy to use and guaranteed to converge. We use the approach to derive optimal control strategies based on COVID-19 data from New Delhi and discuss implications of the optimal control in view of the controlled epidemic trajectories derived from the model.


\section{Mathematical Modelling \label{sec:model}}

We use a classical compartmentalised SEIR model that divides the total population in \emph{susceptible} ($S$), \emph{exposed} ($E$), \emph{infected} ($I$), and \emph{removed} ($R$) individuals \cite{Allen2008,Britton2010}. Susceptible individuals have not been afflicted by the disease and can acquire it; exposed individuals acquired the disease but it is still latent, i.e. it is yet to manifest and become contagious; infected individuals have already manifested the disease and can transmit it; finally, removed individuals have undergone whole disease cycle and can no longer be affected by the condition. At a given day $k \ge 0$, the total number of new contagions, i.e. the increase in the exposed population, is a Poisson variable with rate

\begin{equation} \label{eq:inputrate}
    \lambda(k) = \beta S(k) I(k), \; k \ge 0,
\end{equation}
where $\beta >0$ is the rate of transmission per encounter. In addition, $S(k)$ and $I(k)$ are the numbers of susceptible and infected individuals at the outset of day $k \ge 0$.

Let $\sigma$ be a discrete random variable representing the length of the latency period, in days. Similarly, $\gamma$ is a discrete variable representing the duration of the infectious period, in days. As infections evolve independently and since there is no limit on the number of concomitant infections, the exposition-to-infection phase can be modelled as a $M_t/G/\infty$ queue \cite{Arruda2021novel}. Let $\delta_e(k)$ be the number of new contagions on day $k \ge 0$. It follows from \cite{Eick1993} that $\delta_e(k)$ is a Poisson variable with rate:

\begin{equation} \label{eq:inputinfected}
\overline{\delta}_e(k) =  \mathcal{E}(\lambda  \,( k - \sigma) \,) = \sum_{s = 0}^{M_\sigma} \lambda  \,( k - s) P(\sigma = s),
\end{equation}
where $\Omega_\sigma = \{0, \, \ldots, \, M_\sigma \}$ is the sample space of $\sigma$, $\mathcal E$ is the expected value operator, $\overline{\delta}_e(k) = \mathcal E(\delta_e(k))$ is the expected value of $\delta_e(k)$, and $s \ge 0$ is a discrete time delay. Note that $M_\sigma$ is the maximum possible value assumed by the variable $\sigma$, i.e. the maximum length of the latency period in days. A similar reasoning yields that the infection-to-removal queue is also $M_t/G/\infty$. Therefore, the number of new removals on day $k \ge 0$ is a Poisson variable $\delta_i(k)$ with rate \cite{Eick1993}:

\begin{equation} \label{eq:outputinfected}
\overline{ \delta}_i(k)  = \mathcal E(\delta_e( \,k -  \gamma \,) ) = \sum_{s = 0}^{M_\sigma + M_\gamma} \lambda  \,( k - s) P(\zeta = s),
\end{equation}
where $\zeta = \sigma + \gamma$, $\Omega_\gamma = \{0, \, \ldots, \, M_\gamma\}$ is the sample space of $\gamma$, $\overline{\delta}_i(k) = \mathcal E(\delta_i(k))$ is the expected value of $\delta_i(k)$, and $s \ge 0$ is a discrete time delay. Note that $M_\gamma$ is the maximum possible value assumed by variable $\gamma$, i.e. the maximum length of the infectious period in days.

The proposed model evolves in two levels. The input and output rates in Eq. \eqref{eq:inputrate}, \eqref{eq:inputinfected} and \eqref{eq:outputinfected} vary daily. At each day, the respective rates drive a time-varying stochastic process $X_t, \, t \ge 0$ that evolves in continuous time and monitors each individual jump in the population. $X_t = (S_t, E_t, I_t, R_t), \, t\ge 0$ is a four-dimensional stochastic process that monitors the susceptible ($S_t$), exposed ($E_t$), infected ($I_t$) and removed ($R_t$) populations in time.

Let $\mathcal{N}=\{0,\,1,2,\,3,\, ..., \,N\}$ and $\Omega=\mathcal{N}^4$, where $N$ is a finite non-negative integer and $\Omega$ is the state space of the SEIR populations. Additionally, let $\mathbb Z_+$ be the set of non-negative integers. At any time $t \ge 0$, $X(t) = (\, S(t), \, E(t), \, I(t), R(t) \, ) \in \Omega$ is the state of process $X_t, \, t \ge 0$. By definition, $X_t, \, t \ge 0$ will be subject to random jumps at rate:

\begin{equation} \label{eq:ratelamb}
    \Lambda(k) = \lambda(k) + \overline{\delta}_e(k) + \overline{\delta}_i(k),
\end{equation}
where $k = \lfloor t \rfloor := \max \{s > 0: t \ge s, \; s \in \mathbb{Z_+}\}$ is the integer part of $t$. Let $\{\tau_0, \, \tau_1, \ldots\}$ be the sequence of jumps in the system, with $\tau_0 \equiv 0$ and $\tau_{m+1} > \tau_m, \, \forall m \ge 0$. At time $t = \tau_m$, process $X_t$ jumps to state $X(t^+)$ according to the following probability distribution:%

\begin{multline} \label{eq:transitons}
P(\, X(t^+) = Y | X(t) = (S(t), \, E(t),\, I(t), \, R(t) ), \,t=\tau_m  \,) \\
=\begin{cases} 
\dfrac{\lambda(k)}{\Lambda(k)} & \text{if} \; Y = (S(t) - 1, \; E(t)+1,\; I(t), \; R(t) \, ), \\
\dfrac{\overline{\delta}_e(k)}{\Lambda(k)} & \text{if} \; Y = (S(t) , \; E(t)-1,\; I(t) + 1, \; R(t) \, ), \\
\dfrac{\overline{\delta}_i(k)}{\Lambda(k)} & \text{if} \; Y = (S(t) , \; E(t),\; I(t) - 1, \; R(t) + 1) \, ), \\
0 & \text{otherwise},
\end{cases}
\end{multline}
where $X(t^+)$ is the state of the system just after time $t$. The first expression on the right-hand side of Eq. \eqref{eq:transitons} is the probability of a new contagion; the second represents the manifestation of a previously latent infection and the third corresponds to a new removal.

\begin{figure*}

\tikzset{every picture/.style={line width=0.75pt}} 
\begin{tikzpicture}[x=0.75pt,y=0.75pt,yscale=-1,xscale=1]
\draw  [color={rgb, 255:red, 74; green, 144; blue, 226 }  ,draw opacity=1 ] (14,138) .. controls (14,128.06) and (21.61,120) .. (31,120) .. controls (40.39,120) and (48,128.06) .. (48,138) .. controls (48,147.94) and (40.39,156) .. (31,156) .. controls (21.61,156) and (14,147.94) .. (14,138) -- cycle ;
\draw  [color={rgb, 255:red, 74; green, 144; blue, 226 }  ,draw opacity=1 ] (82,140) .. controls (82,130.06) and (89.61,122) .. (99,122) .. controls (108.39,122) and (116,130.06) .. (116,140) .. controls (116,149.94) and (108.39,158) .. (99,158) .. controls (89.61,158) and (82,149.94) .. (82,140) -- cycle ;
\draw  [color={rgb, 255:red, 74; green, 144; blue, 226 }  ,draw opacity=1 ] (194,138) .. controls (194,128.06) and (203.18,120) .. (214.5,120) .. controls (225.82,120) and (235,128.06) .. (235,138) .. controls (235,147.94) and (225.82,156) .. (214.5,156) .. controls (203.18,156) and (194,147.94) .. (194,138) -- cycle ;
\draw  [color={rgb, 255:red, 74; green, 144; blue, 226 }  ,draw opacity=1 ] (261,138) .. controls (261,128.06) and (268.61,120) .. (278,120) .. controls (287.39,120) and (295,128.06) .. (295,138) .. controls (295,147.94) and (287.39,156) .. (278,156) .. controls (268.61,156) and (261,147.94) .. (261,138) -- cycle ;
\draw  [color={rgb, 255:red, 74; green, 144; blue, 226 }  ,draw opacity=1 ] (327,138) .. controls (327,128.06) and (336.4,120) .. (348,120) .. controls (359.6,120) and (369,128.06) .. (369,138) .. controls (369,147.94) and (359.6,156) .. (348,156) .. controls (336.4,156) and (327,147.94) .. (327,138) -- cycle ;
\draw  [color={rgb, 255:red, 74; green, 144; blue, 226 }  ,draw opacity=1 ] (457,137) .. controls (457,127.06) and (465.73,119) .. (476.5,119) .. controls (487.27,119) and (496,127.06) .. (496,137) .. controls (496,146.94) and (487.27,155) .. (476.5,155) .. controls (465.73,155) and (457,146.94) .. (457,137) -- cycle ;
\draw  [color={rgb, 255:red, 74; green, 144; blue, 226 }  ,draw opacity=1 ] (524,138) .. controls (524,128.06) and (531.61,120) .. (541,120) .. controls (550.39,120) and (558,128.06) .. (558,138) .. controls (558,147.94) and (550.39,156) .. (541,156) .. controls (531.61,156) and (524,147.94) .. (524,138) -- cycle ;
\draw [color={rgb, 255:red, 128; green, 128; blue, 128 }  ,draw opacity=1 ]   (210,118) .. controls (249.6,95.23) and (241.18,82.26) .. (276.9,118.88) ;
\draw [shift={(278,120)}, rotate = 225.76] [color={rgb, 255:red, 128; green, 128; blue, 128 }  ,draw opacity=1 ][line width=0.75]    (10.93,-3.29) .. controls (6.95,-1.4) and (3.31,-0.3) .. (0,0) .. controls (3.31,0.3) and (6.95,1.4) .. (10.93,3.29)   ;
\draw [color={rgb, 255:red, 128; green, 128; blue, 128 }  ,draw opacity=1 ]   (278,120) .. controls (317.6,97.23) and (309.18,84.26) .. (344.9,120.88) ;
\draw [shift={(346,122)}, rotate = 225.76] [color={rgb, 255:red, 128; green, 128; blue, 128 }  ,draw opacity=1 ][line width=0.75]    (10.93,-3.29) .. controls (6.95,-1.4) and (3.31,-0.3) .. (0,0) .. controls (3.31,0.3) and (6.95,1.4) .. (10.93,3.29)   ;
\draw [color={rgb, 255:red, 128; green, 128; blue, 128 }  ,draw opacity=1 ]   (473,118) .. controls (512.6,95.23) and (504.18,82.26) .. (539.9,118.88) ;
\draw [shift={(541,120)}, rotate = 225.76] [color={rgb, 255:red, 128; green, 128; blue, 128 }  ,draw opacity=1 ][line width=0.75]    (10.93,-3.29) .. controls (6.95,-1.4) and (3.31,-0.3) .. (0,0) .. controls (3.31,0.3) and (6.95,1.4) .. (10.93,3.29)   ;
\draw [color={rgb, 255:red, 128; green, 128; blue, 128 }  ,draw opacity=1 ]   (204,154) .. controls (228.63,173.7) and (200.86,177.88) .. (167.53,158.88) ;
\draw [shift={(166,158)}, rotate = 390.47] [color={rgb, 255:red, 128; green, 128; blue, 128 }  ,draw opacity=1 ][line width=0.75]    (10.93,-3.29) .. controls (6.95,-1.4) and (3.31,-0.3) .. (0,0) .. controls (3.31,0.3) and (6.95,1.4) .. (10.93,3.29)   ;
\draw [color={rgb, 255:red, 128; green, 128; blue, 128 }  ,draw opacity=1 ]   (95,162) .. controls (60.52,184.66) and (60.98,184.02) .. (32.33,157.24) ;
\draw [shift={(31,156)}, rotate = 403.03] [color={rgb, 255:red, 128; green, 128; blue, 128 }  ,draw opacity=1 ][line width=0.75]    (10.93,-3.29) .. controls (6.95,-1.4) and (3.31,-0.3) .. (0,0) .. controls (3.31,0.3) and (6.95,1.4) .. (10.93,3.29)   ;
\draw [color={rgb, 255:red, 128; green, 128; blue, 128 }  ,draw opacity=1 ]   (278,156) .. controls (243.52,178.66) and (245.92,182.88) .. (217.33,156.24) ;
\draw [shift={(216,155)}, rotate = 403.03] [color={rgb, 255:red, 128; green, 128; blue, 128 }  ,draw opacity=1 ][line width=0.75]    (10.93,-3.29) .. controls (6.95,-1.4) and (3.31,-0.3) .. (0,0) .. controls (3.31,0.3) and (6.95,1.4) .. (10.93,3.29)   ;
\draw [color={rgb, 255:red, 128; green, 128; blue, 128 }  ,draw opacity=1 ]   (344,156) .. controls (309.52,178.66) and (311.92,182.88) .. (283.33,156.24) ;
\draw [shift={(282,155)}, rotate = 403.03] [color={rgb, 255:red, 128; green, 128; blue, 128 }  ,draw opacity=1 ][line width=0.75]    (10.93,-3.29) .. controls (6.95,-1.4) and (3.31,-0.3) .. (0,0) .. controls (3.31,0.3) and (6.95,1.4) .. (10.93,3.29)   ;
\draw [color={rgb, 255:red, 128; green, 128; blue, 128 }  ,draw opacity=1 ]   (536,156) .. controls (501.53,178.66) and (503.92,182.88) .. (475.33,156.24) ;
\draw [shift={(474,155)}, rotate = 403.03] [color={rgb, 255:red, 128; green, 128; blue, 128 }  ,draw opacity=1 ][line width=0.75]    (10.93,-3.29) .. controls (6.95,-1.4) and (3.31,-0.3) .. (0,0) .. controls (3.31,0.3) and (6.95,1.4) .. (10.93,3.29)   ;
\draw [color={rgb, 255:red, 128; green, 128; blue, 128 }  ,draw opacity=1 ]   (404,157) .. controls (369.7,179.54) and (378.62,179.99) .. (363.93,155.53) ;
\draw [shift={(363,154)}, rotate = 418.39] [color={rgb, 255:red, 128; green, 128; blue, 128 }  ,draw opacity=1 ][line width=0.75]    (10.93,-3.29) .. controls (6.95,-1.4) and (3.31,-0.3) .. (0,0) .. controls (3.31,0.3) and (6.95,1.4) .. (10.93,3.29)   ;
\draw [color={rgb, 255:red, 155; green, 155; blue, 155 }  ,draw opacity=1 ]   (12,129) .. controls (7.05,106.23) and (-1.82,64.84) .. (30.02,118.35) ;
\draw [shift={(31,120)}, rotate = 239.49] [color={rgb, 255:red, 155; green, 155; blue, 155 }  ,draw opacity=1 ][line width=0.75]    (10.93,-3.29) .. controls (6.95,-1.4) and (3.31,-0.3) .. (0,0) .. controls (3.31,0.3) and (6.95,1.4) .. (10.93,3.29)   ;
\draw [color={rgb, 255:red, 128; green, 128; blue, 128 }  ,draw opacity=1 ]   (557,127) .. controls (601.33,105.33) and (593.26,172.92) .. (557.65,152.97) ;
\draw [shift={(556,152)}, rotate = 391.87] [color={rgb, 255:red, 128; green, 128; blue, 128 }  ,draw opacity=1 ][line width=0.75]    (10.93,-3.29) .. controls (6.95,-1.4) and (3.31,-0.3) .. (0,0) .. controls (3.31,0.3) and (6.95,1.4) .. (10.93,3.29)   ;
\draw [color={rgb, 255:red, 128; green, 128; blue, 128 }  ,draw opacity=1 ]   (99,122) .. controls (97.04,95.54) and (108.53,98.85) .. (133.46,108.41) ;
\draw [shift={(135,109)}, rotate = 201.04] [color={rgb, 255:red, 128; green, 128; blue, 128 }  ,draw opacity=1 ][line width=0.75]    (10.93,-3.29) .. controls (6.95,-1.4) and (3.31,-0.3) .. (0,0) .. controls (3.31,0.3) and (6.95,1.4) .. (10.93,3.29)   ;
\draw [color={rgb, 255:red, 128; green, 128; blue, 128 }  ,draw opacity=1 ]   (348,120) .. controls (346.04,93.54) and (357.53,96.85) .. (382.46,106.41) ;
\draw [shift={(384,107)}, rotate = 201.04] [color={rgb, 255:red, 128; green, 128; blue, 128 }  ,draw opacity=1 ][line width=0.75]    (10.93,-3.29) .. controls (6.95,-1.4) and (3.31,-0.3) .. (0,0) .. controls (3.31,0.3) and (6.95,1.4) .. (10.93,3.29)   ;
\draw [color={rgb, 255:red, 128; green, 128; blue, 128 }  ,draw opacity=1 ]   (474,155) .. controls (439.7,177.54) and (438.05,177.99) .. (422.94,153.53) ;
\draw [shift={(422,152)}, rotate = 418.39] [color={rgb, 255:red, 128; green, 128; blue, 128 }  ,draw opacity=1 ][line width=0.75]    (10.93,-3.29) .. controls (6.95,-1.4) and (3.31,-0.3) .. (0,0) .. controls (3.31,0.3) and (6.95,1.4) .. (10.93,3.29)   ;

\draw (273,128) node [anchor=north west][inner sep=0.75pt]   [align=left] {$\displaystyle x$};
\draw (196,128) node [anchor=north west][inner sep=0.75pt]   [align=left] {$\displaystyle x-1$};
\draw (329,128) node [anchor=north west][inner sep=0.75pt]   [align=left] {$\displaystyle x+1$};
\draw (459,127) node [anchor=north west][inner sep=0.75pt]   [align=left] {$\displaystyle x_{m}$-1};
\draw (531,129) node [anchor=north west][inner sep=0.75pt]   [align=left] {$\displaystyle x_{m}$};
\draw (26,130) node [anchor=north west][inner sep=0.75pt]   [align=left] {$\displaystyle 0$};
\draw (94,132) node [anchor=north west][inner sep=0.75pt]   [align=left] {$\displaystyle 1$};
\draw (136,131) node [anchor=north west][inner sep=0.75pt]   [align=left] {$\displaystyle \dotsc $};
\draw (388,132) node [anchor=north west][inner sep=0.75pt]   [align=left] {$\displaystyle \dotsc $};
\draw (6,74) node [anchor=north west][inner sep=0.75pt]  [font=\footnotesize] [align=left] {$\displaystyle 1$};
\draw (85,77) node [anchor=north west][inner sep=0.75pt]  [font=\footnotesize,color={rgb, 255:red, 208; green, 2; blue, 27 }  ,opacity=1 ] [align=left] {$\displaystyle p_{12}^{u}$};
\draw (44,182) node [anchor=north west][inner sep=0.75pt]  [font=\footnotesize,color={rgb, 255:red, 65; green, 117; blue, 5 }  ,opacity=1 ] [align=left] {$\displaystyle p_{10}^{u}$};
\draw (220,77) node [anchor=north west][inner sep=0.75pt]  [font=\footnotesize,color={rgb, 255:red, 208; green, 2; blue, 27 }  ,opacity=1 ] [align=left] {$\displaystyle p_{x-1,x}$};
\draw (590,126) node [anchor=north west][inner sep=0.75pt]  [font=\footnotesize,color={rgb, 255:red, 208; green, 2; blue, 27 }  ,opacity=1 ] [align=left] {$\displaystyle p_{x_{m} ,x_{m}}$};
\draw (168,172) node [anchor=north west][inner sep=0.75pt]  [font=\footnotesize,color={rgb, 255:red, 65; green, 117; blue, 5 }  ,opacity=1 ] [align=left] {$\displaystyle p_{x-1,x-2}^{u}$};
\draw (232,175) node [anchor=north west][inner sep=0.75pt]  [font=\footnotesize,color={rgb, 255:red, 65; green, 117; blue, 5 }  ,opacity=1 ] [align=left] {$\displaystyle p_{x,x-1}^{u}$};
\draw (299,177) node [anchor=north west][inner sep=0.75pt]  [font=\footnotesize,color={rgb, 255:red, 65; green, 117; blue, 5 }  ,opacity=1 ] [align=left] {$\displaystyle p_{x+1,x}^{u}$};
\draw (366,176) node [anchor=north west][inner sep=0.75pt]  [font=\footnotesize,color={rgb, 255:red, 65; green, 117; blue, 5 }  ,opacity=1 ] [align=left] {$\displaystyle p_{x+2,x+1}^{u}$};
\draw (428,174) node [anchor=north west][inner sep=0.75pt]  [font=\footnotesize,color={rgb, 255:red, 65; green, 117; blue, 5 }  ,opacity=1 ] [align=left] {$\displaystyle p_{x_{m} -1,x_{m} -2}^{u}$};
\draw (492,176) node [anchor=north west][inner sep=0.75pt]  [font=\footnotesize,color={rgb, 255:red, 65; green, 117; blue, 5 }  ,opacity=1 ] [align=left] {$\displaystyle p_{x_{m} ,x_{m} -1}^{u}$};
\draw (299,79.33) node [anchor=north west][inner sep=0.75pt]  [font=\footnotesize,color={rgb, 255:red, 208; green, 2; blue, 27 }  ,opacity=1 ] [align=left] {$\displaystyle p_{x,x+1}$};
\draw (349,80.33) node [anchor=north west][inner sep=0.75pt]  [font=\footnotesize,color={rgb, 255:red, 208; green, 2; blue, 27 }  ,opacity=1 ] [align=left] {$\displaystyle p_{x+1,x+2}$};
\draw (486,75.33) node [anchor=north west][inner sep=0.75pt]  [font=\footnotesize,color={rgb, 255:red, 208; green, 2; blue, 27 }  ,opacity=1 ] [align=left] {$\displaystyle p_{x_{m -1},x_{m}}$};
\end{tikzpicture}
\caption{Controlled random-walk dynamics \label{fig:controllequeue}}
\end{figure*}
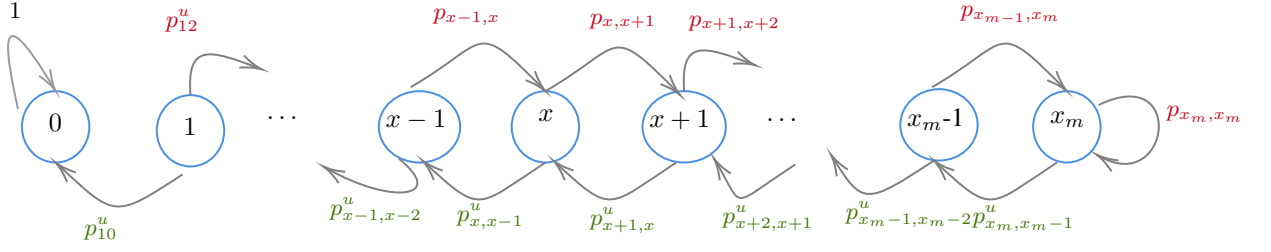

\subsection{Controlled dynamics \label{sec:contdyn}}

To control the epidemic's spread, assume that on day $k \ge 0$ we apply a mitigation $0 \le u(k) \le 1$ to prevent new contagions. The mitigation prevents a proportion $u(k)$ of new transmissions and produces a new controlled rate of contagion:
\begin{equation} \label{eq:cont_infection}
    \lambda(k, u(k)) = \beta (1 - u(k) ) S(k) I(k).
\end{equation}
This affects the dynamics of the system. Let $U$ be a discrete set of feasible mitigating actions in the interval $[0,1)$ and let $\pi = \{u(k) \in U, \, k \ge 0\}$ be a feasible mitigation policy. Then, the epidemic is characterized by a controlled Markov chain $X_t, \, t \ge 0$. Under the controlled dynamics, we have:
\begin{equation} \label{eq:inputinfected_controlled}
\overline{\delta}_e^\pi(t) = \sum_{w = 0}^{M_\sigma} \lambda  \,(k, u(k - w)) P(\sigma = w),
\end{equation}
\begin{equation} \label{eq:outputinfected_controlled}
\overline{ \delta}_i^\pi(t) = \sum_{w = 0}^{M_\sigma + M_\gamma} \lambda  \,( k - w) P(\zeta = w),
\end{equation}
and
\begin{equation} \label{eq:ratelamb_controlled}
    \Lambda^\pi(t) = \lambda(k, u(k)) + \overline{\delta}_e^\pi(k) + \overline{\delta}_i^\pi(k),
\end{equation}
with $k = \lfloor t \rfloor$ representing the number of days elapsed since the epidemic's outset, and $w$ represents a positive time delay.

Let $\{\tau_0, \, \tau_1, \ldots\}$ be the sequence of jump times of process $X_t, \, t \ge 0$. Then, assuming $k = \lfloor t \rfloor$ at a given jump time $t \ge 0$, we have the controlled probability distribution $P^{\pi}$ as: 

\begin{scriptsize}
\begin{multline} \label{eq:cont_transitons}
P^\pi(\, X(t^+) = Y | X(t) = (S(t), \, E(t),\, I(t), \, R(t), \, \tau = \tau_m \,) \\
=\begin{cases} 
\dfrac{\lambda(k, u(k))}{\Lambda^\pi(k)} & \text{if} \; Y = (S(t) - 1, \; E(t)+1,\; I(t), \; R(t) \, ), \\
\dfrac{\overline{\delta}_e^\pi(k)}{\Lambda^\pi(k)} & \text{if} \; Y = (S(t) , \; E(t)-1,\; I(t) + 1, \; R(t) \, ), \\
\dfrac{\overline{\delta}_i^\pi(k)}{\Lambda^\pi(k)} & \text{if} \; Y = (S(t) , \; E(t),\; I(t) - 1, \; R(t) + 1) \, ), \\
0 & \text{otherwise}.
\end{cases}
\end{multline}
\end{scriptsize}
%

\subsection{Near-optimal strategies}

To control the dynamics in Eq. \eqref{eq:cont_transitons}, we need the information of the input rates for the whole disease cycle - the past $M_\sigma + M_\gamma$ days, as in Eq. \eqref{eq:inputinfected_controlled}-\eqref{eq:outputinfected_controlled}. For practical purposes, however, this often results in an intractable Markov decision process. For example, if the maximum disease cycle is $7$ days, the state of the Markov decision process would be $S = \mathcal R^7_+ \cup \Omega$, where the seven positive continuous variables represent the transmission-rate history over the last 7 days.

Let us now consider the following hypotheses:
\begin{hyp} \label{hyp:beds}
$\overline{ \delta}_i^\pi(t) \le \bar \rho \lambda(k, u(k))$ whenever $E(k) + I(k) \ge m T_p$, where $T_p$ is the total population at $t=k$, $0 < m < 1$ and $0 < \bar \rho < 1$;
\end{hyp}
\begin{hyp} \label{hyp:stability}
$S(t) >>> E(t) + I(t)$.
\end{hyp}

Hypothesis \ref{hyp:beds} is due to the limited number of hospital beds in most countries \cite{OECD2018,Tarrataca2021}, and additional access barriers in some settings \cite{Goldwasser2016,Angelo2017}. It ensures that effective mitigation is enforced when healthcare systems are close to nominal capacity. Effective mitigation imposes a contagion rate that is always surpassed by the removal rate. The recent COVID-19 outbreak demonstrated that healthcare systems around the globe tend to collapse in the early stages of an epidemic with  moderate hospitalization rate \cite{Ramachandran2020,Lemos2020}. Hence, the limited healthcare capacity will force early mitigation (Hypothesis \ref{hyp:stability}, see also \cite{Tarrataca2021}) and limit the decrease of susceptible population. This motivates Hypothesis \ref{hyp:stability}.

Considering Hypotheses \ref{hyp:beds} and \ref{hyp:stability}, we introduce a simplified Markov decision process to control the evolution of the exposed and infected populations. The controlled process $Z_k, \, k \ge 0$ is discrete and evolves in state space $S= \{0, \, 1, \, \ldots, \, z_m\}$, that represents the total number of infections and expositions at any given period $k \ge 0$; $z_m$ denotes the maximum allowed number of expositions and infections. Let $\bar \sigma = \mathcal E(\sigma)$ denote the expected values of the latency period and infectious periods and $\bar \gamma = \mathcal E(\gamma)$ be the expected value of the infectious period. For each state-action pair $(i, u)$ in $S \times U$, the one-step transition probabilities $p_{ij}^u$ for all $j \in S$ are defined as:

\begin{equation} \label{eq:transmdp}
p_{ij}^u = \begin{cases}
\dfrac{R_0 (1-u) i}{\bar \sigma + \bar \gamma} \cfrac{1}{W}, & \text{if} \; j = \min(i+1, z_m), \\
\dfrac{i}{\bar \sigma +  \bar \gamma} \dfrac{1}{W}, & \text{if} \; j = i-1,\\
1 - p_{i(i+1)} - p_{i(i-1)}, & \text{if} \; i = j, \\
0, & \text{otherwise},
\end{cases}
\end{equation}
where $R_0 =  \beta S \bar \gamma$ and 
\[W =  \dfrac{R_0 z_m}{\bar \sigma + \bar \gamma} + \dfrac{z_m}{\bar \sigma +  \bar \gamma}
\]
is a normalizing factor, to allow each transition to occur within the same expected time interval.
Fig. \ref{fig:controllequeue} illustrates the transition probabilities for a fixed $u \in U$. The first expression on the right-hand side of Eq. \eqref{eq:transmdp} is the application of Eq. \eqref{eq:cont_infection}, with $I(k) = \dfrac{\bar \gamma}{\bar \sigma} E(k)$ and $i = E(k)+I(k)$, subjected to normalization factor $W$. This approximation is based on Little's law \cite{Shortle2018}, that yields:
\[
E(k) = \bar{\lambda} \bar \sigma, \quad I(k) = \bar{\lambda} \bar \gamma,
\]
where $\bar \lambda$ is the average rate of contagion. The second expression in the right-hand side of Eq. \eqref{eq:transmdp} is the normalized removal rate of the current infected population, equivalent to $\dfrac{I(k)}{\bar \gamma} \dfrac{1}{W}$.

At each step, the system incurs a cost $c: S \times U \to \mathcal R_+$, depending on the current state-action pair $(i,u)$:
\begin{equation}\label{eq:cost}
c(i, u ) = c_1 i + e^{c_2 (i - \bar z)} + e^{c_3 u},
\end{equation}

where $0 < \bar z < z_m$ is a proxy for the target bed occupation level, $c_1, \, c_2, \, c_3$ are positive scalar parameters. Let $\pi: S \to U$ be a stationary closed-loop control policy as defined in Section \ref{sec:contdyn} and let $\Pi$ be the set of all feasible stationary closed-loop control policies. The cost-to-go $V^{\pi}$ of policy $\pi \in \Pi$ is given by:

\begin{equation} \label{eq:value}
    V^\pi(i) = \mathcal{E} \left[ \, \sum_{k=0}^\infty c( Z_k, \pi(Z_k) ) | Z_0 = i \, \right], \; i \in S.
\end{equation}
The objective is to find an optimal policy $\pi^* \in \Pi$ such that:
\begin{equation} \label{eq:optimalpolicy}
    \pi^* = \argmin_{\pi \in \Pi} V^\pi(i), \quad V^*(i) = V^{\pi^*}(i), \; \forall i \in S.
\end{equation}

Since the state space is limited and the cost function $c(\cdot)$ is positive and bounded  there exists an stationary policy \cite{Puterman1994} that satisfies Eq. \eqref{eq:optimalpolicy}. Let $\mathcal V$ be the space of real-valued functions in $S$. To find such a policy, we use the classical value iteration Algorithm \ref{alg:DecomposedEvaluation}, that has guaranteed convergence under the present settings \cite{Puterman1994}.

\begin{algorithm}[ht]
	\KwIn{An arbitrary initial solution $V_0 \in \mathcal V$ and an arbitrary tolerance $\epsilon$}
	\KwOut{Optimal solution $V^{*}$, and optimal stationary policy $\pi^*$ }
	$k \gets 1$\;
	$V_k \gets \infty$\;
	\While{$\| V_k - V_{k-1} \|  \geq \epsilon \,$}{
		\For{each $i \in S$}{
		    \begin{equation}
T V_k(i) := \min_{u \in U} \left\lbrace c(i,u) + \sum_{j \in S} p_{ij}^u V_k(j) \right\rbrace \;
\end{equation}
		}
		\For{each $i \in S$}{
		    \begin{equation}
V_{k+1}(i) := TV_k(i) \;
\end{equation}
		}
		$k \gets  k+1$\;
	}
	$V^* \gets V_k$\;
	$\pi^*(i) \gets \arg \displaystyle \min_{u \in U} \left[ TV^*(i) \right], \forall i \in S$\;
	\Return $V^{*}, \pi^*$.
	\caption{Value iteration}
	\label{alg:DecomposedEvaluation}
\end{algorithm}

\section{Experimental Results \label{sec:experiments}}

We model the experiments for the COVID-19 situation in New Delhi, the capital of India. Considering very low testing in the initial stages of the pandemic and unawareness of the disease, $R_0$ varies. We take $R_0$ as given in \cite{S.MariMuthu2020}. Since $R_0$ varies with time and is higher during initial stages of the pandemic, we conduct the experiments using  $R_0\;=\{2.5,3.5\}$. The Julia and R language source code is available \href{https://github.com/Tarun-Sharma9168/Optimal_Control_And_Decision_Making}{here}.

The parameters of the model are given in Table \ref{tab:Parameters}. Table \ref{tab:Initial Parameters for Stochastic Model} depicts the simulation parameters as per the statistics in \href{https://populationstat.com/india/delhi}{https://populationstat.com/india/delhi}.
\begin{table}[h!]
\centering
\caption{Parameters of the Model}
\label{tab:Parameters}
\begin{tabular}{|c|c|c|}
    \hline  
    Parameters & Description & Unit \\
     \hline\hline
     $\beta$ & Transmission Rate & Transmissions/encounter \\
     $\gamma$ & Recovery Period  & days \\
     $\sigma$ & Latency  Period  & days \\
     \hline\hline
\end{tabular}
\end{table}

\begin{table}[h!]
\centering
\caption{Initial Parameters for Stochastic Model}
\label{tab:Initial Parameters for Stochastic Model}
\begin{tabular}{|c|c|c|}
    \hline  
    Parameters & Description & Value \\
     \hline\hline
     $P$ & Total Population & $31181000$ \\
     $I(0)$ & Initial Infected Population  & $100$ \\
     $E(0)$ & Initial Exposed Population  & $1260$ \\
     $S(0)$ & Initial Susceptible Population & $31181000\;-\;1360$ \\
     $\beta$ & Transmission rate & $R_0/(\gamma*S)$\\
     \hline\hline
\end{tabular}
\end{table}

We assume the distribution of the latency and the recovery periods ($\sigma$ and $\gamma$, respectively) to be discrete to cope with the typical daily collection of the data. Tables \ref{tab:Latency Period Distribution} and \ref{tab:Recovery Period Distribution} show the distributions of the latency and recovery periods, respectively. Table \ref{tab:Parameters for Cost function used in optimal control} introduces the cost parameters used in the experiments as in Eq. \eqref{eq:cost}.

\begin{table}[h!]
\centering
\caption{Latency Period Distribution (maximum period is $14$ days.}
\label{tab:Latency Period Distribution}
\begin{tabular}{|c|c|c|c|c|c|c|}
     \hline\hline
     Day & 0 & 1 & 2 & 3 & 4 & 5\\
     \hline
     Prob & 0.0000 & 0.0009 & 0.0056 & 0.0222 & 0.0611 & 0.1222\\
     \hline\hline
     Day & 6 & 7 & 8 & 9 & 10 & 11\\
     \hline
     Prob & 0.1833 & 0.2095 & 0.1833 & 0.1222 & 0.0611 & 0.0222\\
     \hline\hline
     Day & 12 & 13 & 14 & & & \\
     \hline
     Prob & 0.0056 & 0.0009 & 0.0001 & & &\\
     \hline\hline
\end{tabular}
\end{table}

\begin{table}[h!]
\centering
\caption{Recovery Period Distribution maximum period is $35$ days.}
\label{tab:Recovery Period Distribution}
\begin{tabular}{|c|c|c|c|c|c|c|}
     \hline\hline
     Day & 0 & 1 & 2 & 3 & 4 & 5\\
     \hline
     Prob & 0.0000 & 0.0000 & 0.0000 & 0.0000 & 0.0000 & 0.0000 \\
     \hline\hline
     Day & 6 & 7 & 8 & 9 & 10 & 11\\
     \hline
     Prob & 0.0000 & 0.0000 & 0.0000 & 0.0000 & 0.0000 & 0.0000 \\
     \hline\hline
     Day & 12 & 13 & 14 & 15 & 16 & 17 \\
     \hline
     Prob & 0.0000 & 0.0002 & 0.0010 & 0.0034 & 0.0098 & 0.0233 \\
     \hline\hline
     Day & 18 & 19 & 20 & 21 & 22 & 23 \\ 
     \hline
     Prob & 0.0466 & 0.0792 & 0.1151 & 0.1439 & 0.1550 & 0.1439 \\
     \hline\hline
     Day & 24 & 25 & 26 & 27 & 28 & 29 \\
     \hline
     Prob & 0.1151 & 0.0792 & 0.0466 & 0.0233 & 0.0098 & 0.0034 \\
     \hline\hline
     Day & 30 & 31 & 32 & 33 & 34 & 35 \\
     \hline
     Prob & 0.0010 & 0.0002 & 0.0000 & 0.0000 & 0.0000 & 0.0000\\
     \hline\hline
\end{tabular}
\end{table}

\begin{table}[h!]
\centering
\caption{Parameters for Cost function $c(i,u)$}
\label{tab:Parameters for Cost function used in optimal control}
\begin{tabular}{|c|c|c|c|c|c|}
     \hline\hline
     Parameter Name & c1 & c2 & c3 & z & threshold \\
     \hline
     Parameter Value & 1 & 1 & 1 & 0.015*31181000 & 0.015 \\
     \hline\hline
\end{tabular}
\end{table}

We conduct the experiments with no control measures for $R_0\;=2.5$ and $R_0\;=3.5$, then we introduce control measures that change after every single, $7$, $14$, and $28$ days. 

\begin{figure}[H]
\centering
\includegraphics[width=8cm]{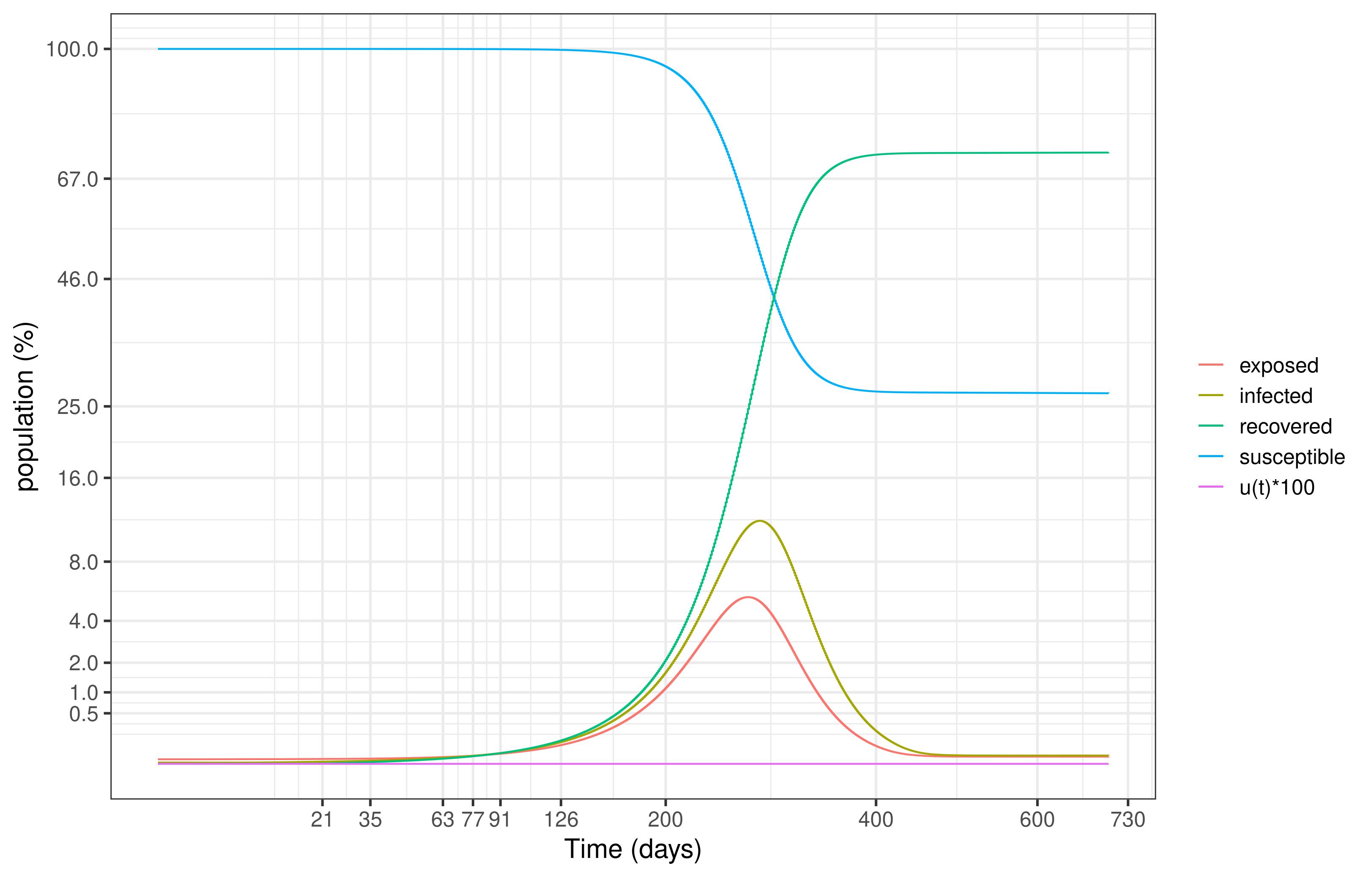}
\caption{The figure shows the daily evolution of the pandemic in the absence of control measures when the reproduction number $R_0=2.5$. The number of individuals in each SEIR block is shown as a percentage of the total population. On the $300^{th}$ day, the infected population reaches approx $12\%$ and the exposed population approx $6\%$. The pandemic stabilises afterwards.}
\label{fig:SEIR plot (R0=2.5 without control)}
\end{figure}

We study the system's behaviour without control measures and for different values of $R_0$, for a period of two years from the pandemic's outset. Fig. \ref{fig:SEIR plot (R0=2.5 without control)} depicts the output of the stochastic model without control measures for $R_0\;=2.5$. On the $300^{th}$ day, the infected population reaches approx $12\%$ and the exposed population approx $6\%$ of the total population, thereafter the pandemic stabilises. By applying some control measures, we can reduce the peak of expositions and infections. Such measures will also stabilise the pandemic earlier than it would otherwise happen without control measures. We use Fig. \ref{fig:SEIR plot (R0=2.5 without control)} as a benchmark to evaluate the effect of the optimal control.

\begin{figure}[H]
\centering
\includegraphics[width=8cm]{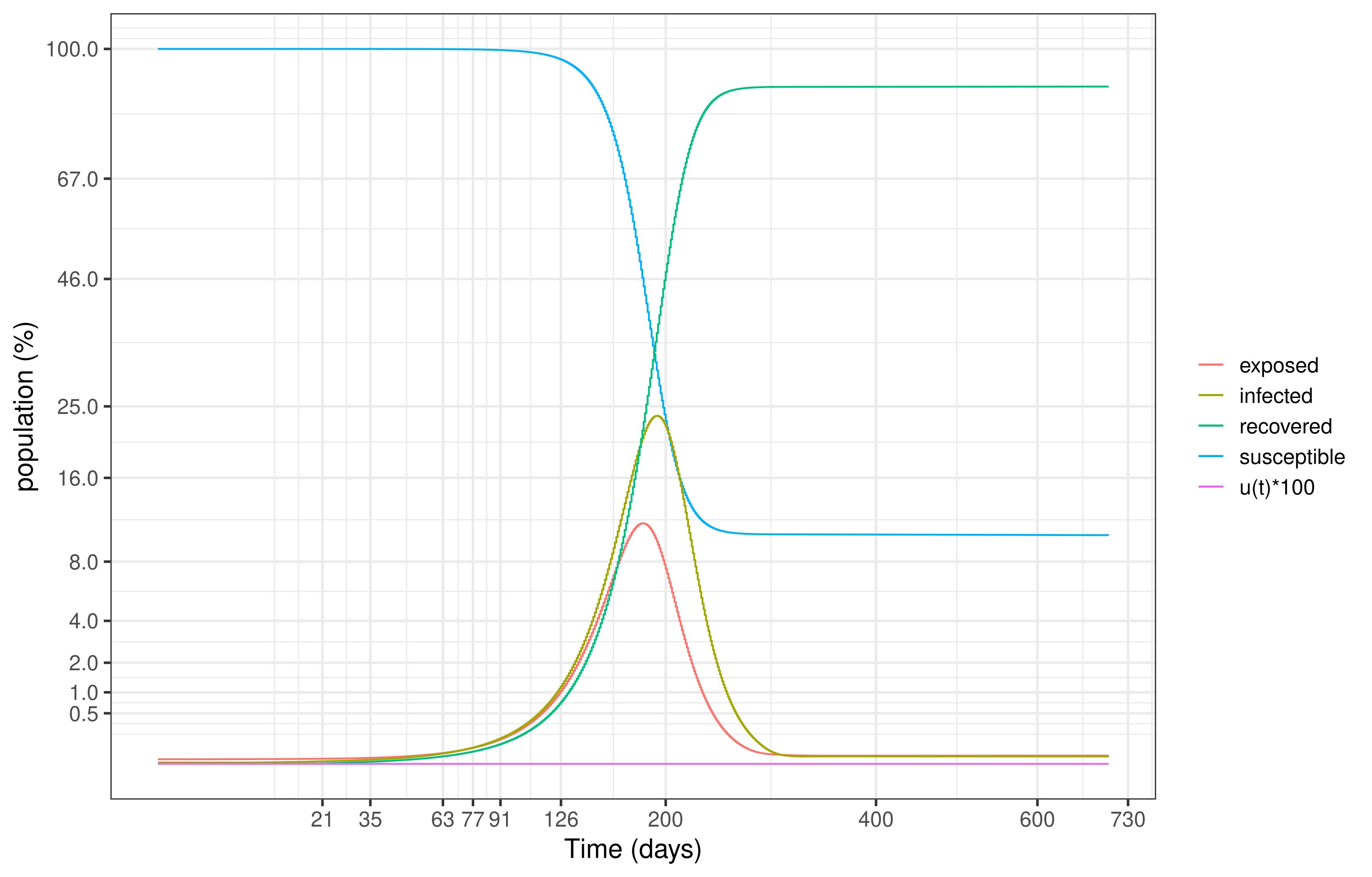}
\caption{ The figure shows the daily evolution of the pandemic in the absence of control measures when the reproduction number $R_0=3.5$. The number of individuals in each SEIR block is shown as a percentage of the total population. On the $180^{th}$ day, the infected population reaches approx $24\%$ and the exposed population approx $13\%$. The pandemic stabilises afterwards.}
\label{fig:SEIR plot (R0=3.5 without control)}
\end{figure}

Fig. \ref{fig:SEIR plot (R0=3.5 without control)} shows the output of the stochastic model with $R_0\;=3.5$ and all remaining parameters unaltered from the previous experiment. As $R_0$ increases, we observe higher peaks for the exposed and infected populations. The former reaches approximately 13\%, whereas the latter peaks at 24\% of the total population. We conclude that higher values of $R_0$ lead to an earlier stabilisation of the pandemic as the cases accumulate faster. For both values of $R_0$, it is clear that the high level of infection will challenge the healthcare resources and demand mitigating measures when these resources become insufficient.

We consider applying control measures to some extent. We assume in all the test cases that the control measures are not $100\%$ effective in order to facilitate the essential services, accommodate variation in individual behaviours, and limit the effect on the economy. We assume that the maximum achievable control (mitigation) effect is $90\%$. The experiments can support policy makers by indicating suitable and timely control measures.

We propose Algorithm \ref{alg:DecomposedEvaluation} to find an optimal control strategy for each $R_0$ and apply this strategy in the original controlled model in Section \ref{sec:contdyn} to assess its effect on the epidemic. We assume that the controls are applied only after the sum of infected and exposed individuals reach $0.1\%$; this is to simulate the delay in the epidemic detection. Below we find the optimal controlled dynamics for different cases. For each experiment, we simulated the stochastic system 100 times and plotted the mean values.

\begin{figure}[H]
\centering
\includegraphics[width=8cm]{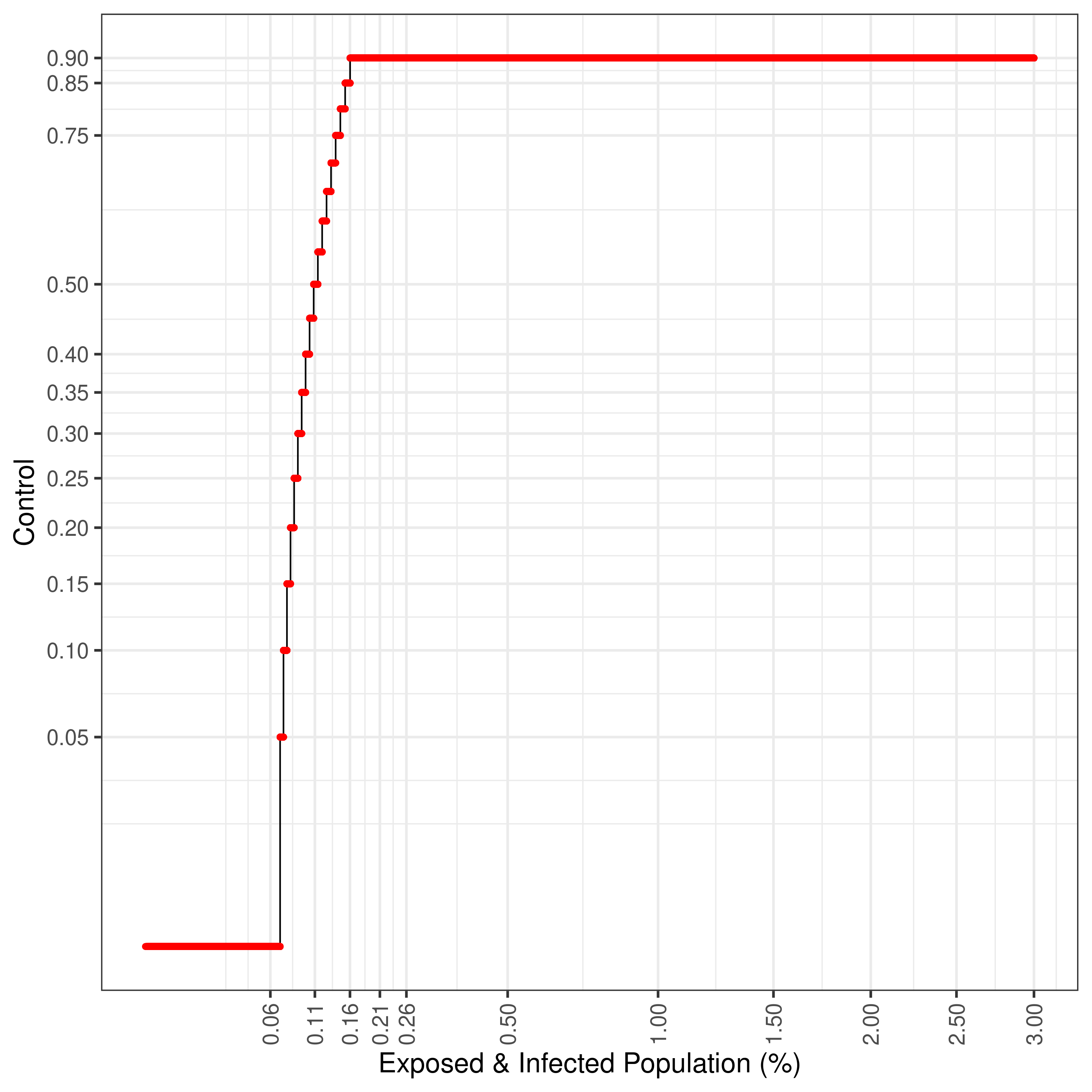}
\caption{Control plot ($R_0\;=2.5$ control vs Population Plot)}
\label{fig:Control plot (R0=2.5 control revised every day)}
\end{figure}

\begin{figure}[H]
\centering
\includegraphics[width=8cm]{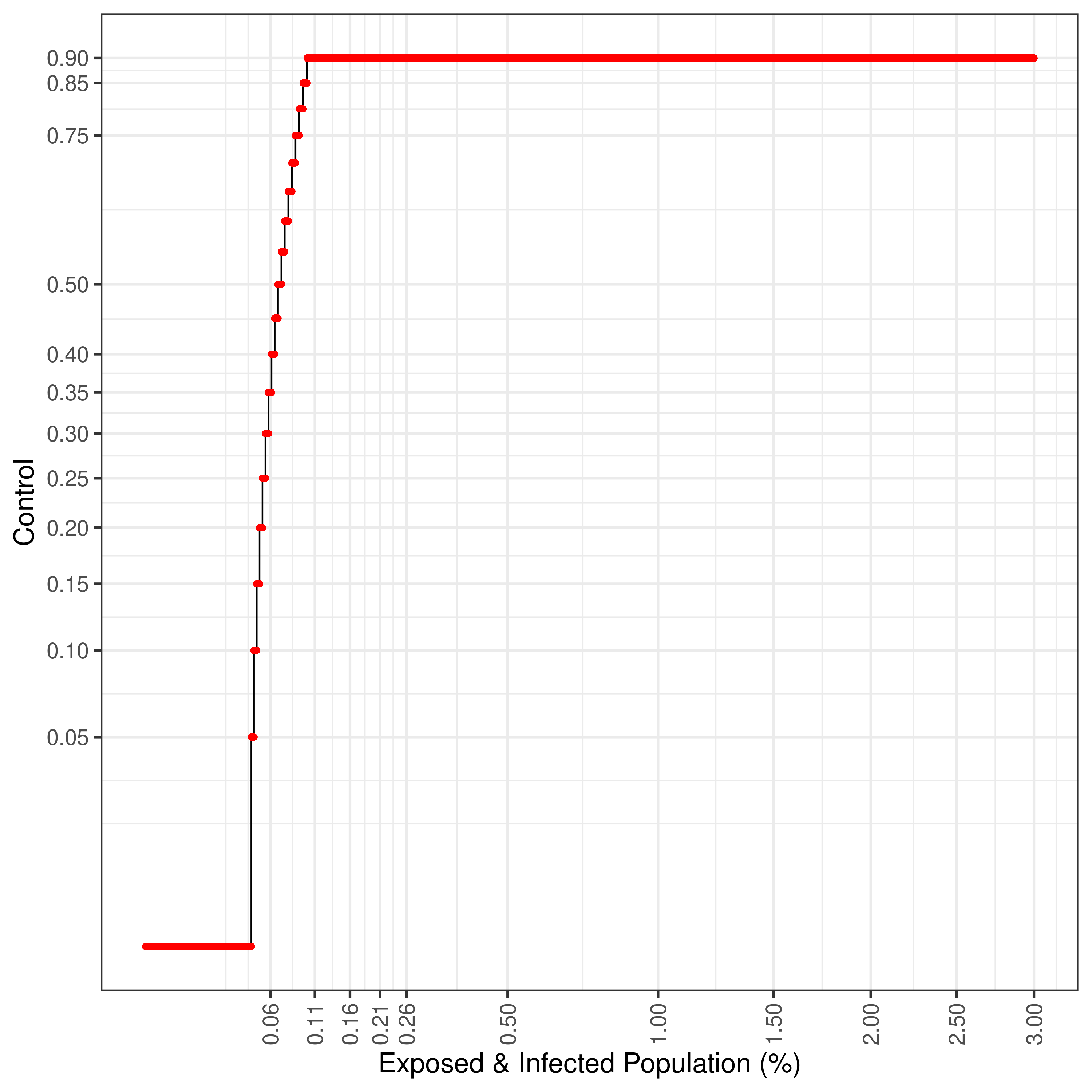}
\caption{Control plot ($R_0\;=3.5$ control vs Population Plot)}
\label{fig:Control plot (R0=3.5 control revised every day)}
\end{figure}

We have the sum of exposed and infected population vs control plot for each $R_0$ in Fig. \ref{fig:Control plot (R0=2.5 control revised every day)} and \ref{fig:Control plot (R0=3.5 control revised every day)}, which depict the approximate control.


\subsection{Experiments with $R_0 = 2.5$}
\begin{figure}[H]
\centering
\includegraphics[width=8cm]{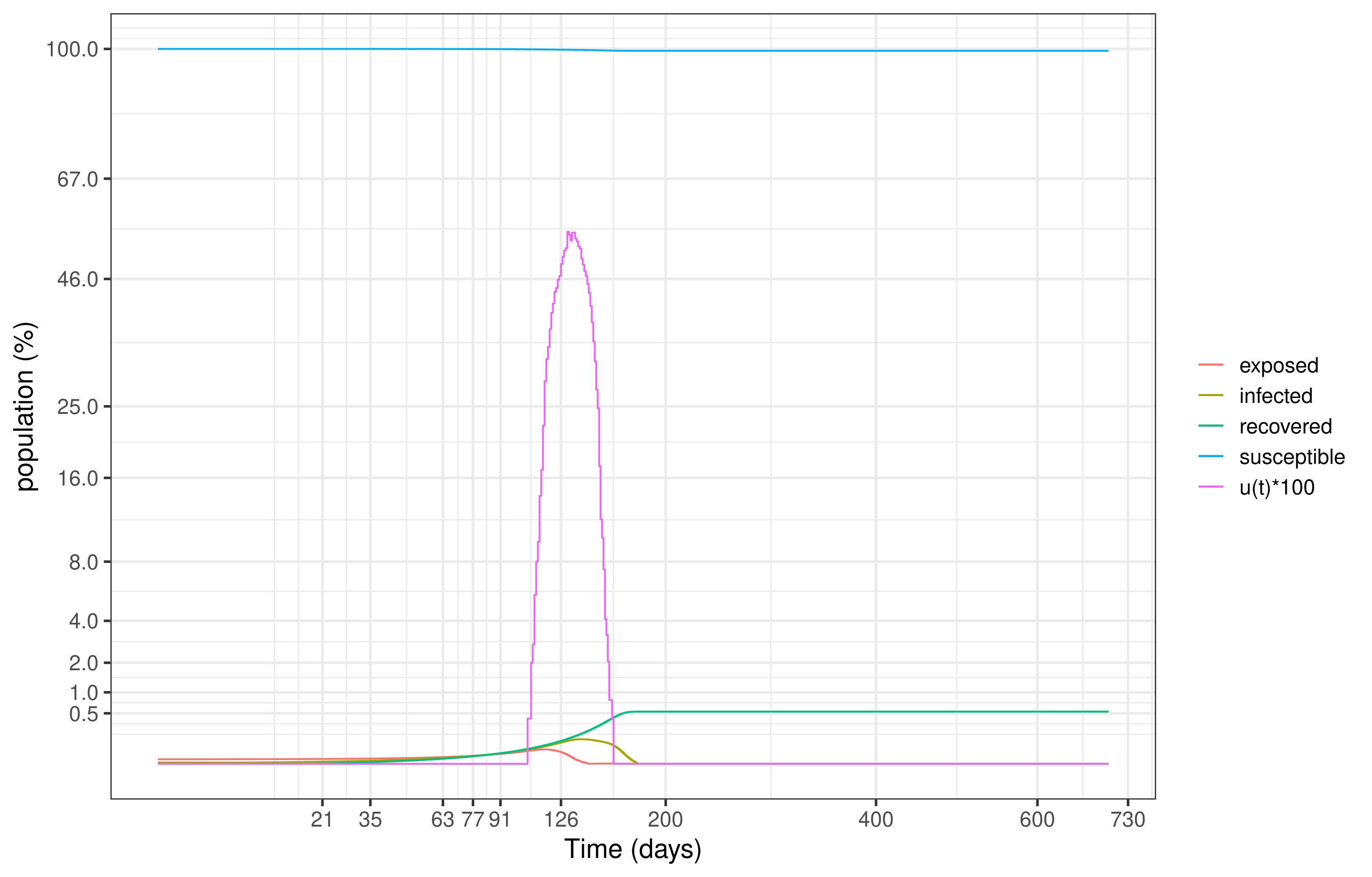}
\caption{SEIR plot ($R_0\;=2.5$ control revised every day)}
\label{fig:SEIR plot (R0=2.5 control revised every day)}
\end{figure}

\begin{figure}[h]
\centering
\includegraphics[width=8cm]{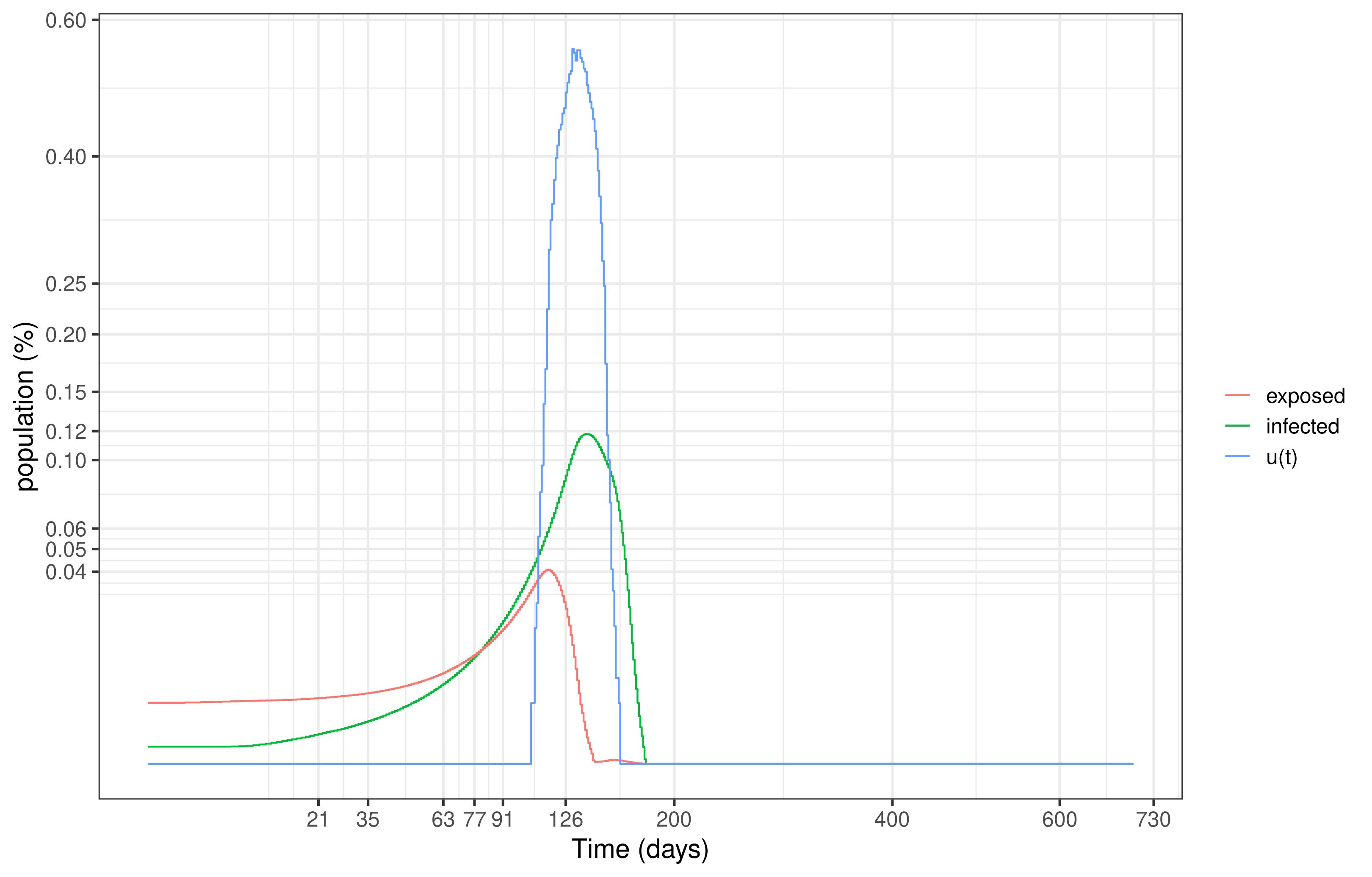}
\caption{Exposed-Infected plot ($R_0\;=2.5$ control revised every day)}
\label{fig:Exposed-Infected plot (R0=2.5 control revised every day)}
\end{figure}

We start the experiments with $R_0\;=2.5$. Fig. \ref{fig:SEIR plot (R0=2.5 control revised every day)} depicts all four population counts with control measures revised daily for Case 1. We observe significant decreases in both the infected and exposed populations. Fig. \ref{fig:Exposed-Infected plot (R0=2.5 control revised every day)} enhances the result for the $E$ and $I$ populations. The control measures reduce the peak of infections to only $0.12\%$ and decrease the peak of expositions to less than $0.05\%$.
This case is considered to be the baseline, but changing control measures daily may not be feasible in practice.

\begin{figure}[h]
\centering
\includegraphics[width=8cm]{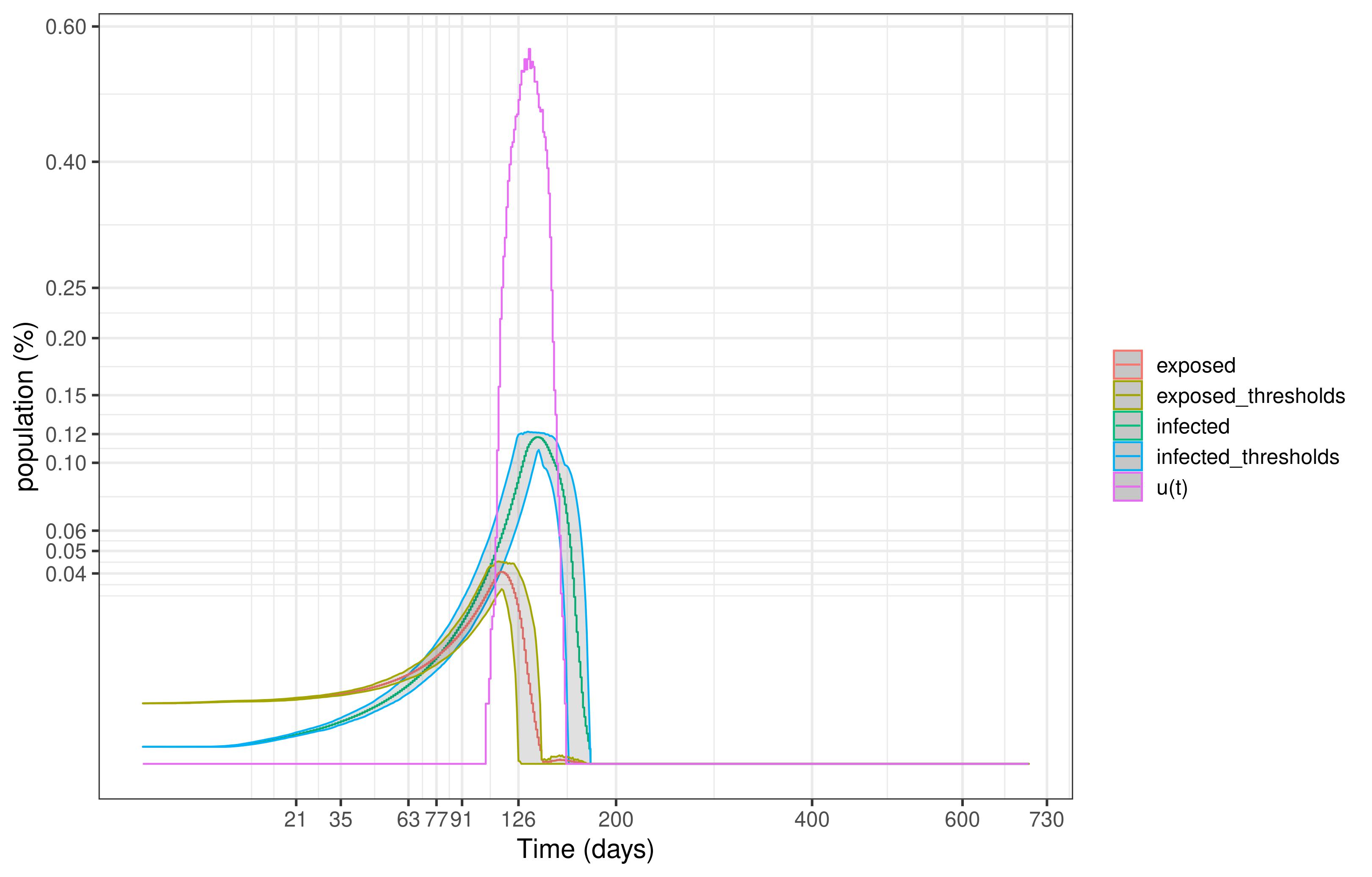}
\caption{The figure shows the confidence interval (CI) for both infected and exposed populations of the pandemic when the control measures are revised daily with reproduction number $R_0=2.5$. }
\label{fig:CI E-I plot (R0=2.5 control revised every day)}
\end{figure}

Fig. \ref{fig:CI E-I plot (R0=2.5 control revised every day)} shows the confidence interval (CI) plot for both infected and exposed populations over $100$ simulations of the proposed stochastic model. We observe that the plot is smooth with no visible spikes due to the regular control changes. 

\begin{figure}[h]
\centering
\includegraphics[width=8cm]{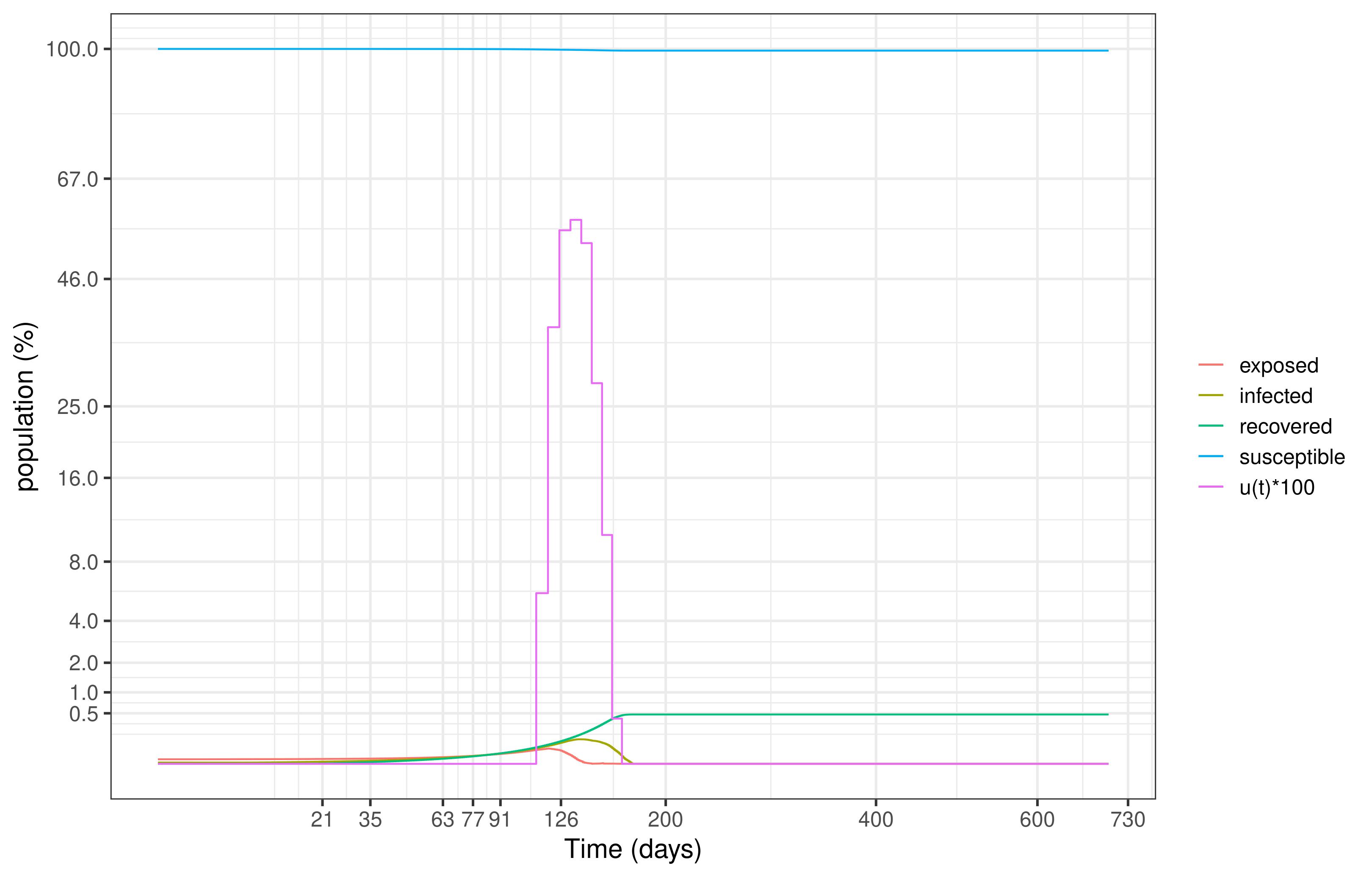}
\caption{SEIR plot ($R_0\;=2.5$ control revised every $7$ days)}
\label{fig:SEIR plot (R0=2.5 control revised every 7 day)}
\end{figure}

\begin{figure}[h]
\centering
\includegraphics[width=8cm]{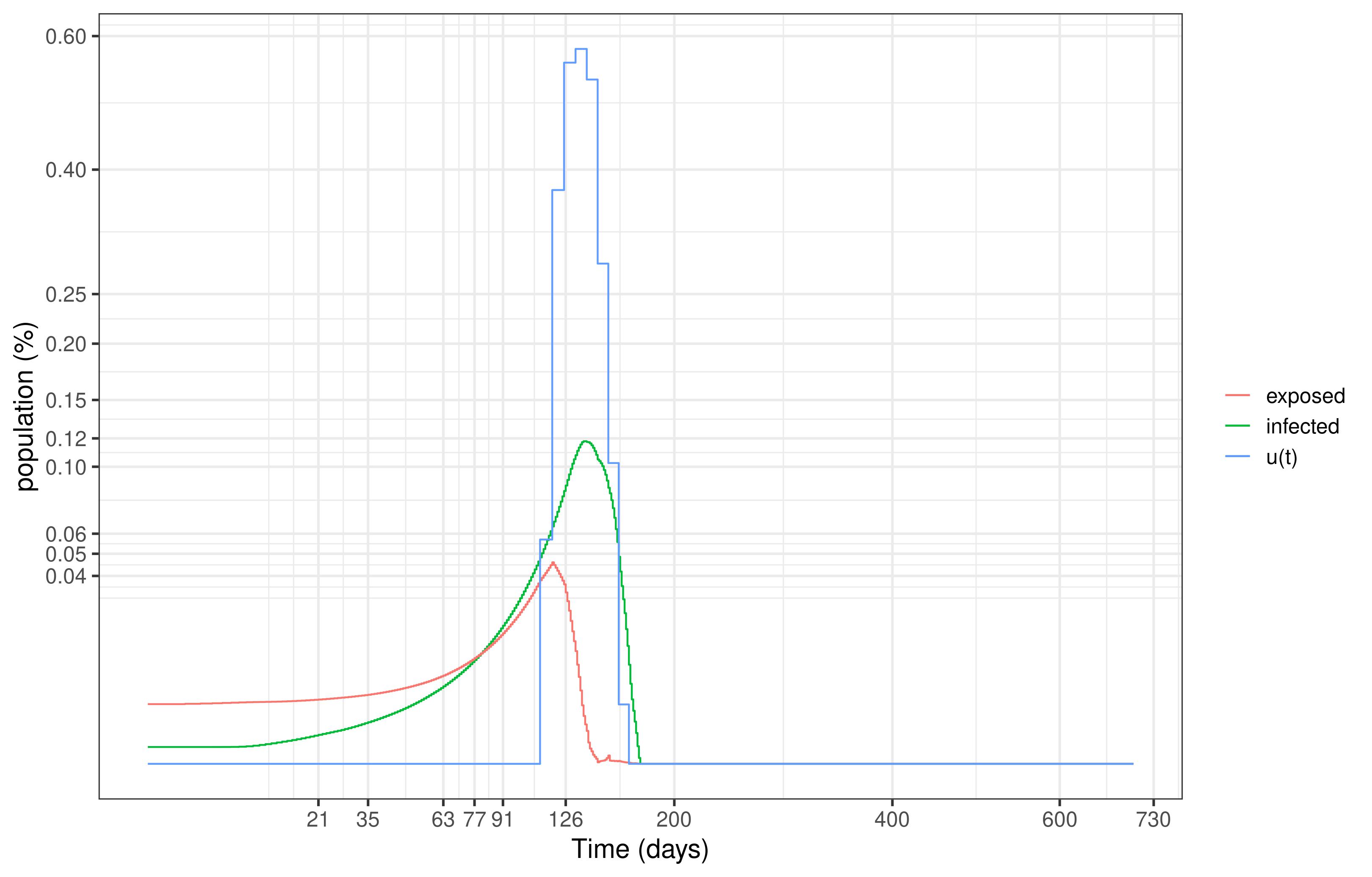}
\caption{Exposed-Infected plot ($R_0\;=2.5$ control revised every $7$ days)}
\label{fig:Exposed-Infected plot(R0=2.5 control revised every 7 day)}
\end{figure}

In Case $2$ we revise the control measures after every $7$ days. We expect an increase in infections and expositions due to the delay in control change; higher control levels may be needed in the middle of the 7-day interval, but the control will not be revised before the next 7-day period. Fig. \ref{fig:SEIR plot (R0=2.5 control revised every 7 day)} shows the SEIR plot for $R_0\;=2.5$ and control measures changing every $7$ days. We observe in Fig. \ref{fig:Exposed-Infected plot(R0=2.5 control revised every 7 day)} that the maximum infected population remains virtually unchanged, but the peak of the exposed population increases slightly with respect to Case 1. 
It is noteworthy that the sum of exposed and infected individuals increases slightly if we delay the change in control for 7 days. 

\begin{figure}[h]
\centering
\includegraphics[width=8cm]{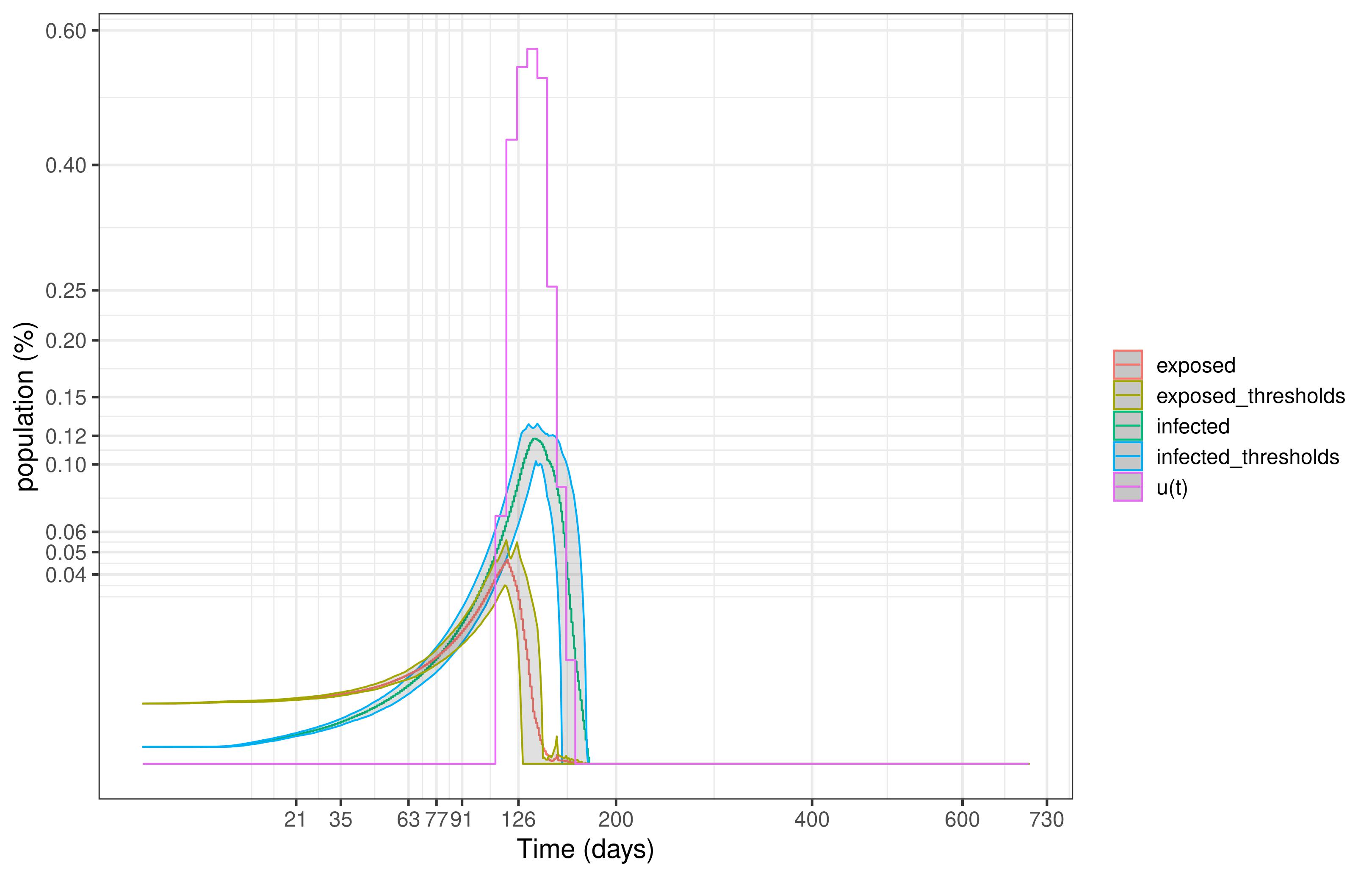}
\caption{CI E-I plot ($R_0\;=2.5$ control revised every 7 days)}
\label{fig:CI E-I plot (R0=2.5 control revised every 7 day)}
\end{figure}

Fig. \ref{fig:CI E-I plot (R0=2.5 control revised every 7 day)} presents the CI for the infected and exposed populations, as well as the optimal control. We observe spikes in the control trajectory, which are necessary to stabilise the system after a 7-day interval in which sub-optimal actions were taken. We need to adjust for overshooting and/or undershooting in the previous interval every $7$ days, leading to spikes in the control and population trajectories. It is noteworthy, however, that the control is able to stabilise the system despite the delayed response to change.

\begin{figure}[h]
\centering
\includegraphics[width=8cm]{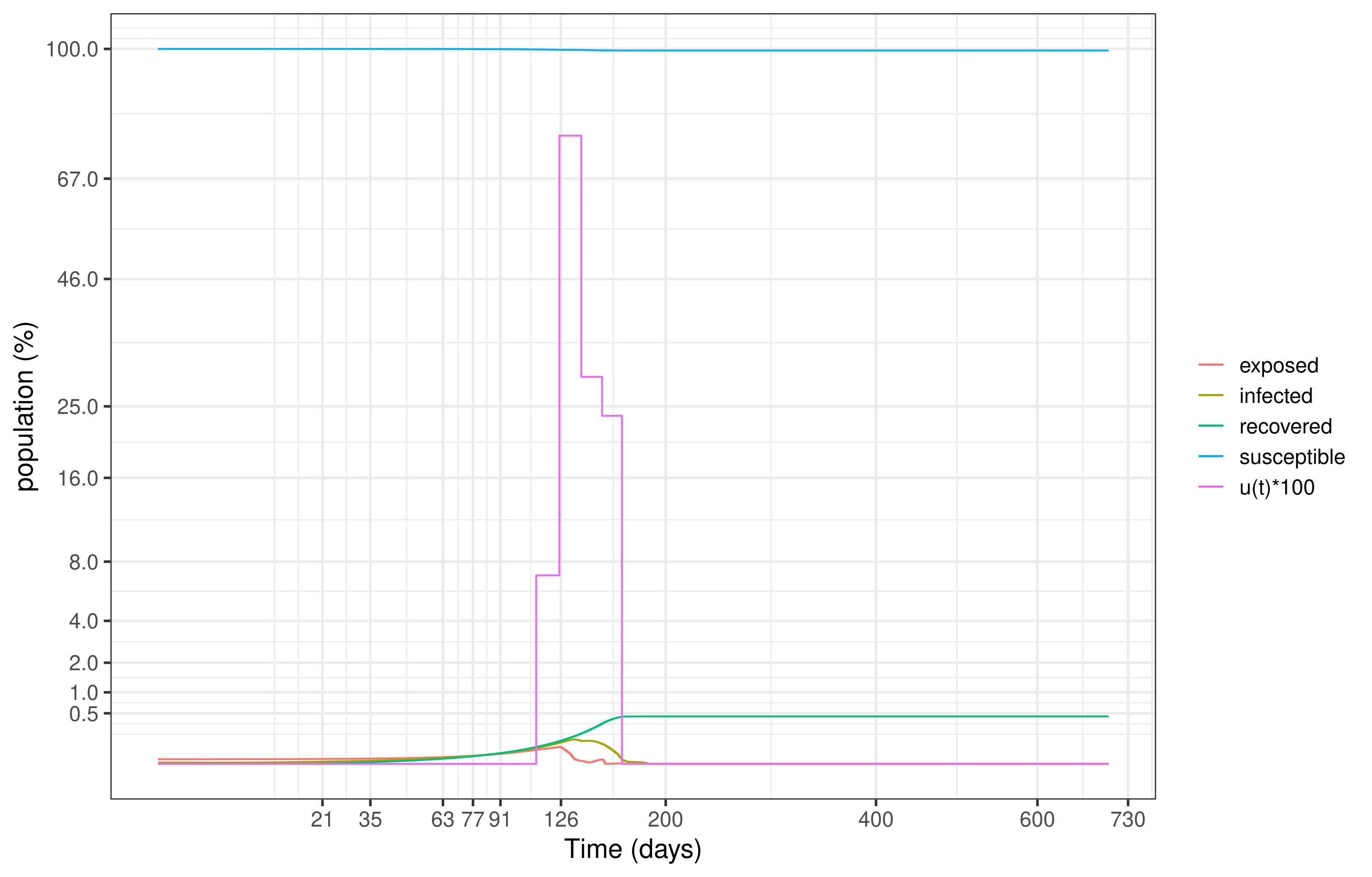}
\caption{SEIR plot ($R_0\;=2.5$ control revised every $14$ days)}
\label{fig:SEIR plot (R0=2.5 control revised every 14 day)}
\end{figure}
\begin{figure}[h]
\centering
\includegraphics[width=8cm]{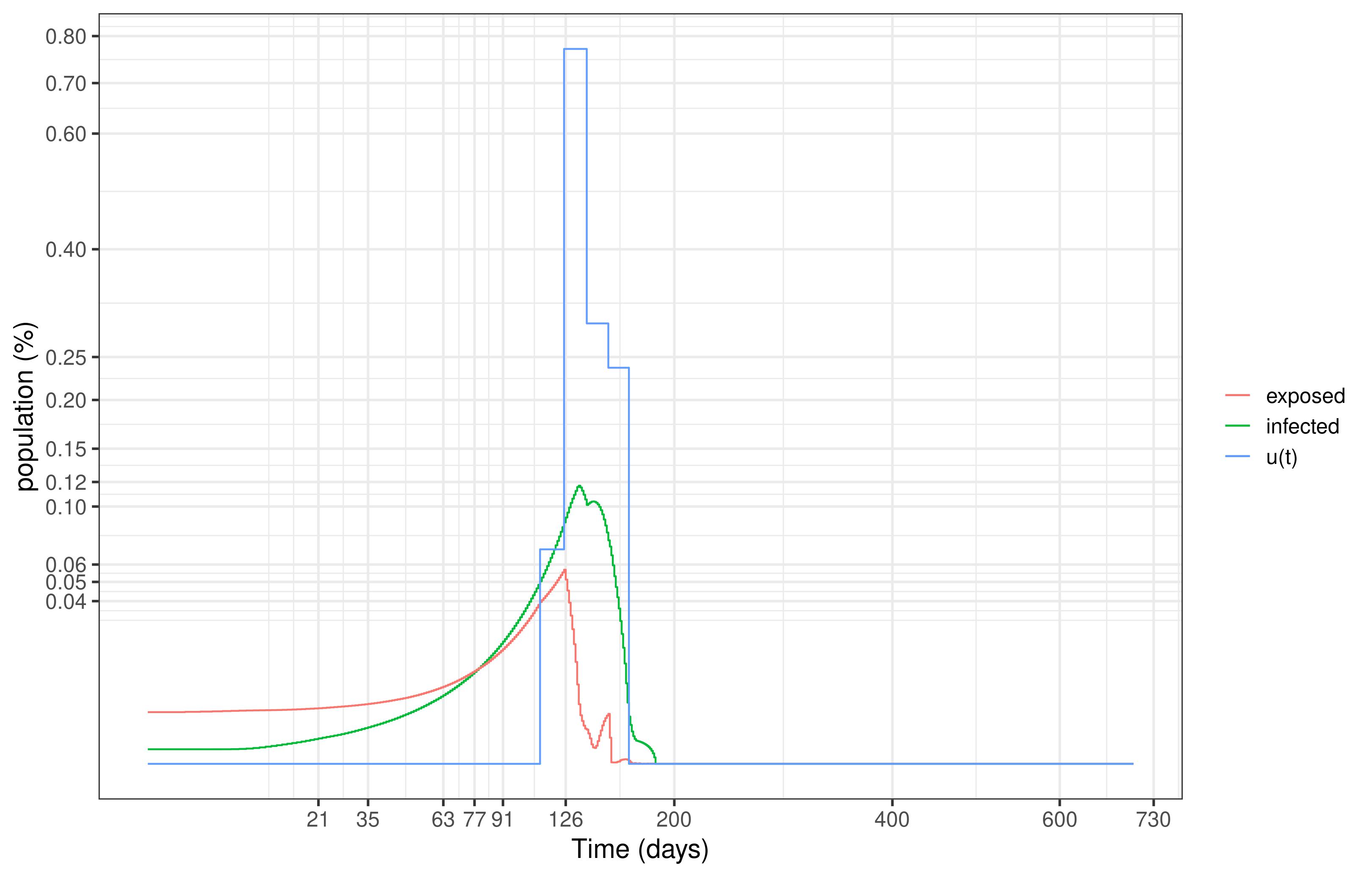}
\caption{Exposed-Infected plot ($R_0\;=2.5$ control revised after $14$ days)}
\label{fig:Exposed-Infected plot(R0=2.5 control revised after 14 days}
\end{figure}
\begin{figure}[h]
\centering
\includegraphics[width=8cm]{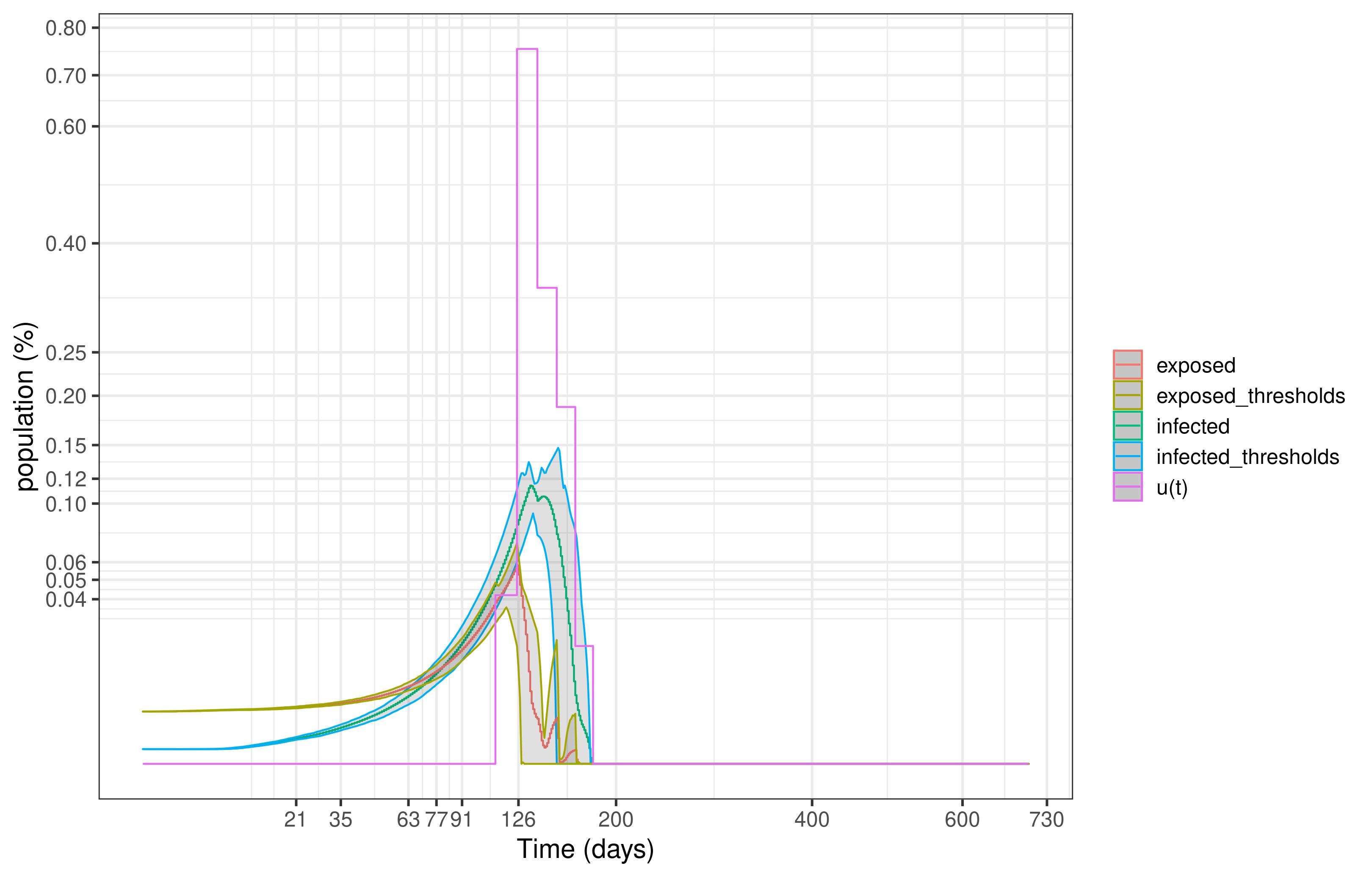}
\caption{CI E-I plot ($R_0\;=2.5$ control revised every 14 days)}
\label{fig:CI E-I plot (R0=2.5 control revised every 14 day)}
\end{figure}

Fig. \ref{fig:SEIR plot (R0=2.5 control revised every 14 day)} shows the output of Case $3$ when $R_0\;=2.5$ with control measures changing every $14$ days. As in the previous case, we expect a slight increase in the exposed and infected populations. Fig. \ref{fig:Exposed-Infected plot(R0=2.5 control revised after 14 days} shows a slight increase in the exposed and infected populations with respect to Case 2. 
Fig. \ref{fig:CI E-I plot (R0=2.5 control revised every 14 day)}  shows the CI plot for the exposed and  infected populations for Case 3, changing controls every fortnight. We notice the increased fluctuations and spikes due to control corrections performed every other week to compensate for sub-optimal actions in the previous fortnight. We observe that the upper bound for the infected population reached $0.15\%$, whereas that value remained close to $0.15\%$ in Case 2.

After $91$ days, the infected population is very low and accordingly control takes a low value. As we can only change the control every fortnight, the infected population increases exponentially. If by the time the control is revised, the infected population has increased too steeply, the control will be high and will lead to a swift decrease in infection. Otherwise, the control level will administer the infection levels more smoothly. As a matter of fact, we will have sufficient time to monitor the effect of the control and this may also decrease the value of epidemic monitoring. This case is very useful in some of the countries where everyday or weekly monitoring is difficult and expensive to perform. The confidence interval chart for the Delhi population shows that we can reduce the infected population with this approach efficiently.

\begin{figure}[H]
\centering
\includegraphics[width=8cm]{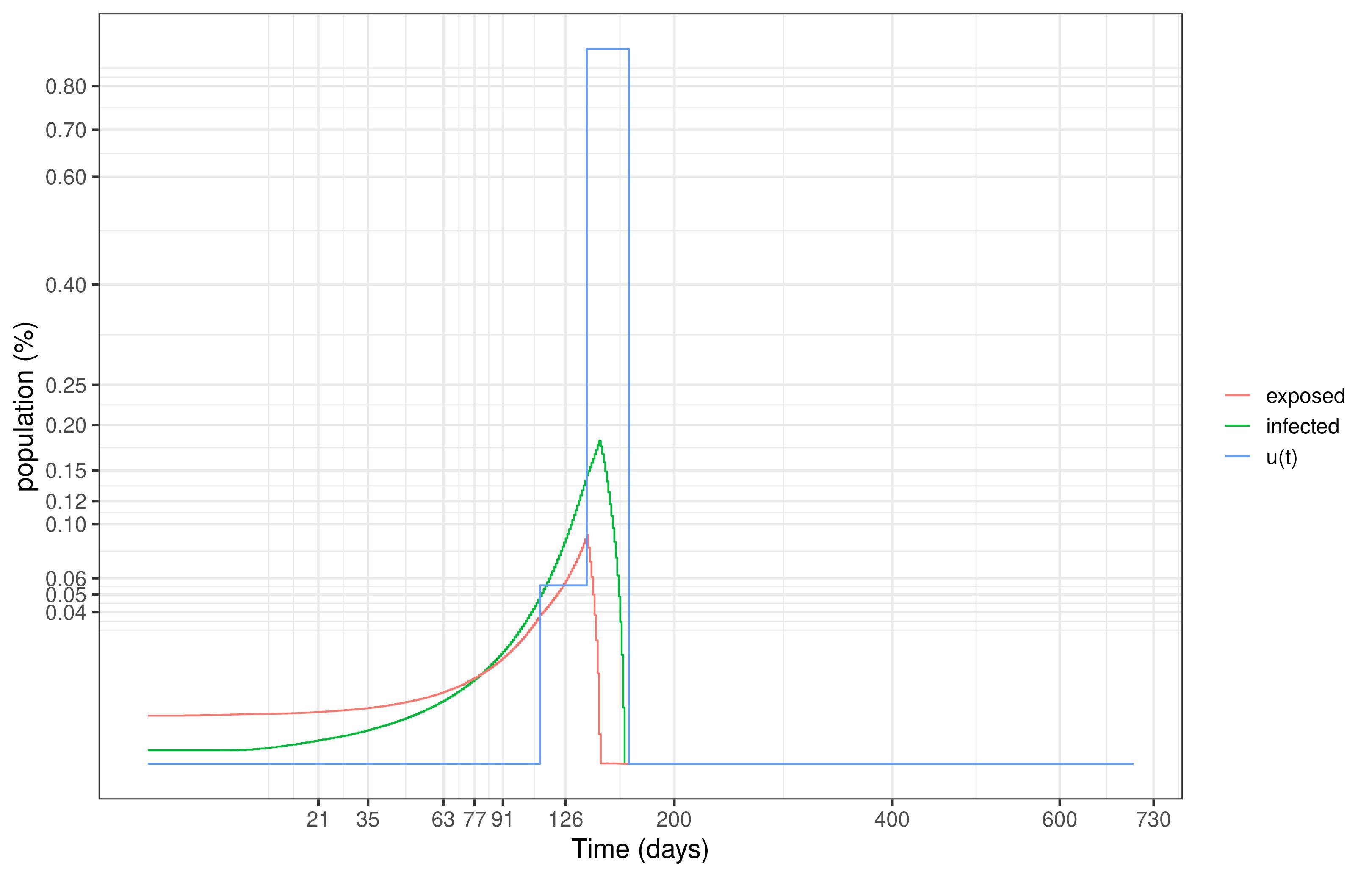}
\caption{Exposed-Infected plot ($R_0\;=2.5$ control revised after $28$ days)}
\label{fig:Exposed-Infected plot(R0=2.5 control revised after 28 days}
\end{figure}

Fig. \ref{fig:Exposed-Infected plot(R0=2.5 control revised after 28 days} depicts the results for Case 4, when the control measures are revised every 4 weeks. We observe that the delay in policy adjustment causes a significant increase in the peak of infections, which ramp up to approximately $0.18\%$ in contrast to the $0.12\%$ peak with daily control adjustment. The exposed population peaks at about $0.09\%$. This experiment has a practical interest because most developed countries use similar intervals to review their epidemic mitigation measures whilst monitoring the progress of the disease. In developing countries, however, policy adjustments are often unplanned and monitoring is not systematic. Furthermore, poor communication may hinder policy changes, as the population may be unaware of the current guidelines.

Fig. \ref{fig:Exposed-Infected plot(R0=2.5 control revised after 28 days} illustrates the effect of large control review intervals. Since it will take $28$ days to react to fluctuations, we expect a more erratic behaviour. Because the optimal control tends to increase with time in the epidemic outset, a large wait may lead to high levels of infection until the control is revised. With low $R_0$ this can be problematic, since the optimal control levels tend to be smaller in the outset of the epidemic and may lead to insufficient mitigation before the next adjustment. That will lead to higher control levels in the next adjustment, as illustrated by the steep increase in control observed after four and eight weeks. This situation is similar to that of India, where low control levels were kept without review for a long time, which resulted in a powerful second wave of infections that required much higher control levels. The results illustrate the importance of properly monitoring the outbreak and adjusting the control levels as often as possible to avoid fluctuations and unintended increase in infections, which may lead to the collapse of the healthcare system.

\begin{figure}[h]
\centering
\includegraphics[width=8cm]{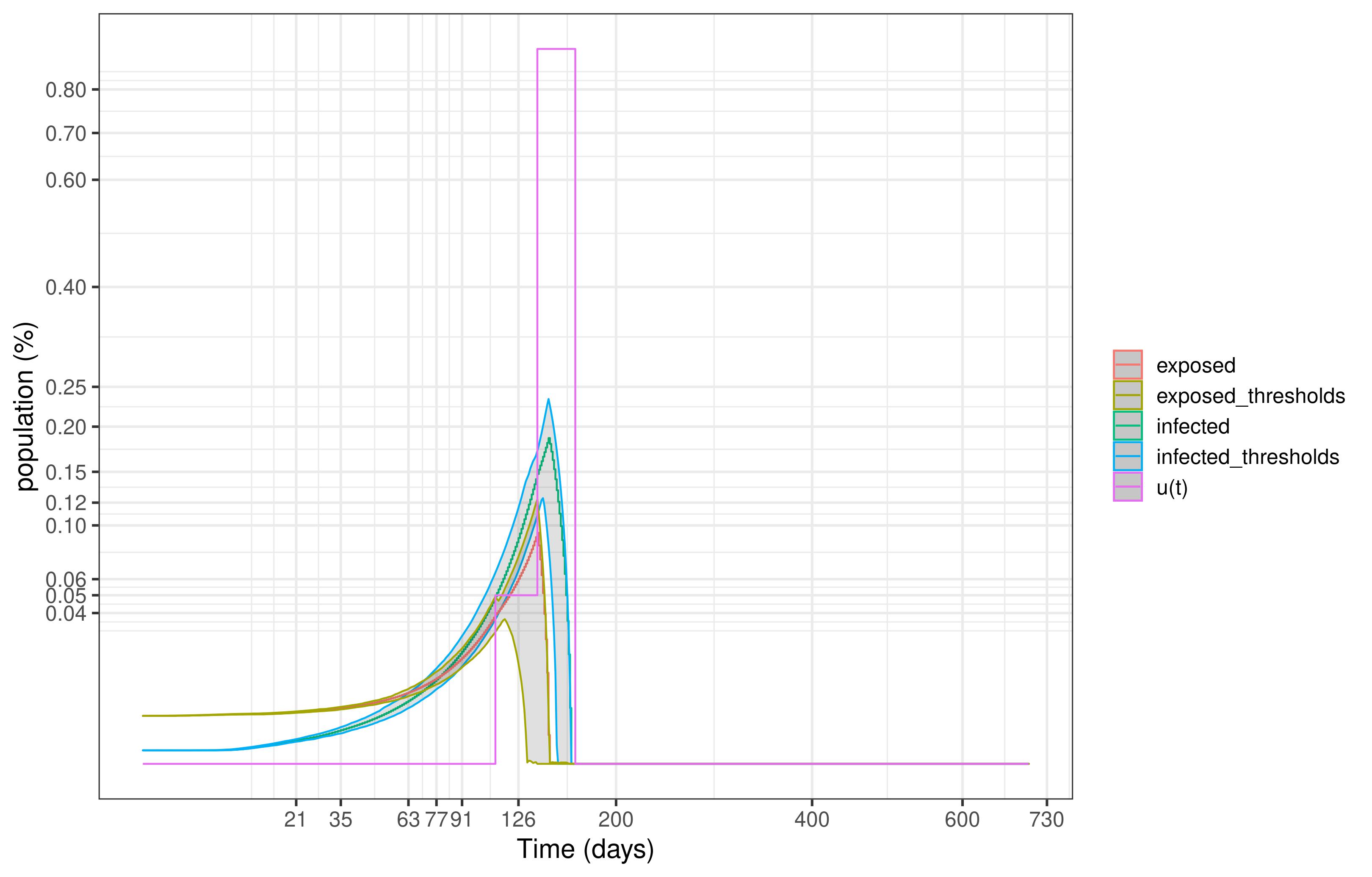}
\caption{CI E-I plot ($R_0\;=2.5$ control revised every 28 days)}
\label{fig:CI E-I plot (R0=2.5 control revised every 28 day)}
\end{figure}

Fig. \ref{fig:CI E-I plot (R0=2.5 control revised every 28 day)} shows the CI for the exposed and infected populations in Case 4. It shows that over a $28-$day interval with initially low control levels, the pandemic has spread and reached an upper bound of $\approx$ $0.25\%$ infected individuals. Hence, we are forced to keep high control levels over a large time interval.

\subsection{Experiments with $R_0 = 3.5$}

This section covers the set of experiments for $R_0\;=3.5$. We notice that a large value of $R_0$ warrants high control levels from the outset, as it is more difficult to control a highly contagious outbreak. A byproduct of this is that the mitigation becomes less sensitive to the control review period.

\begin{figure}[H]
\centering
\includegraphics[width=8cm]{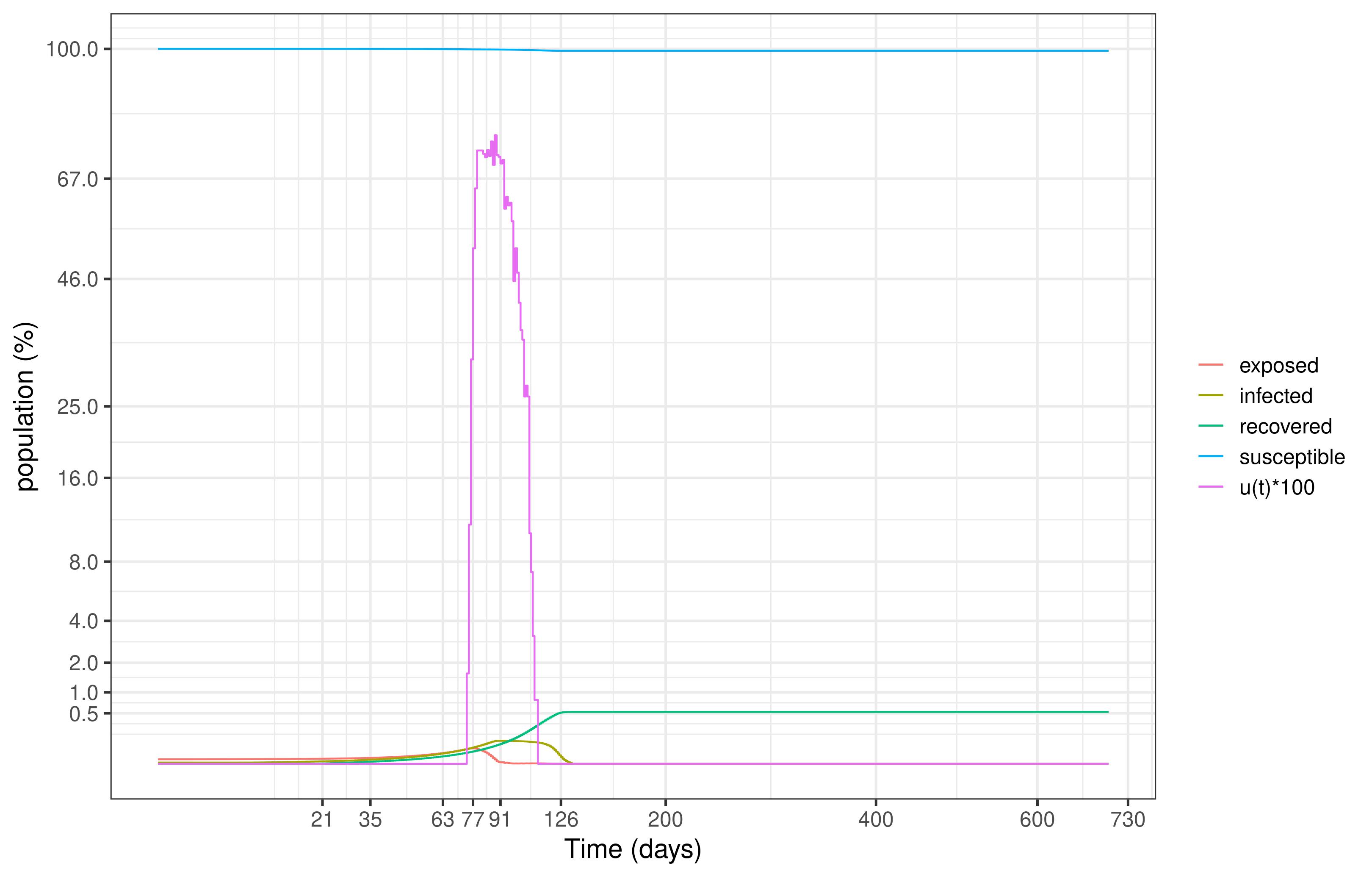}
\caption{SEIR plot ($R_0\;=3.5$ control revised every day)}
\label{fig:SEIR plot(R0=3.5 control revised every days}
\end{figure}

\begin{figure}[h]
\centering
\includegraphics[width=8cm]{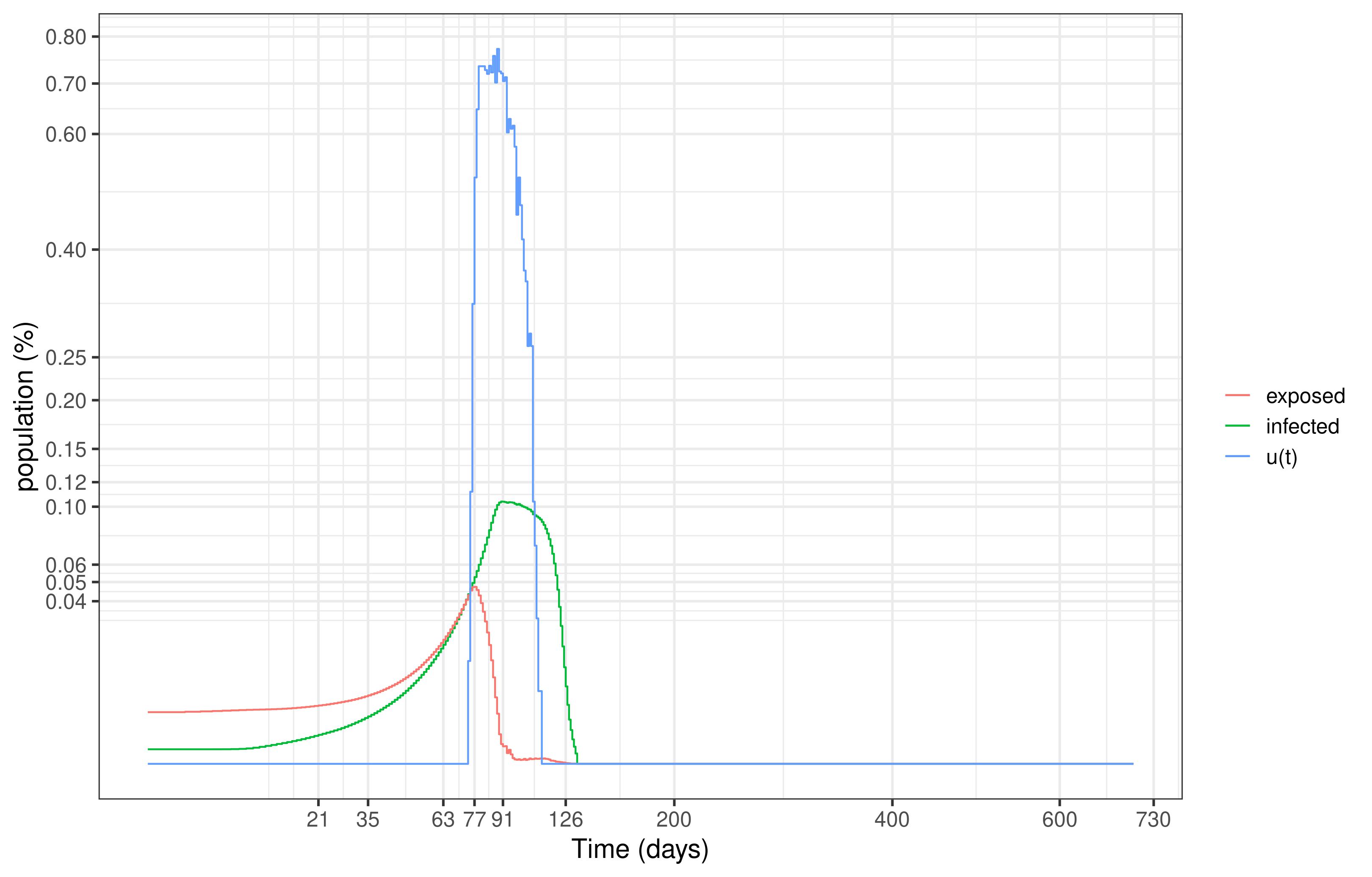}
\caption{Exposed-Infected plot ($R_0\;=3.5$ control revised every day)}
\label{fig:Exposed-Infected plot(R0=3.5 control revised every days}
\end{figure}

Fig. \ref{fig:SEIR plot(R0=3.5 control revised every days} depicts the baseline, the controlled trajectory with control measures revised daily and $R_0\;=3.5$ - Case 1. We see in Fig. \ref{fig:Exposed-Infected plot(R0=3.5 control revised every days} that the optimal strategy effectively curbs the outbreak. The infected population peaks at about $0.11\%$ while the exposed population reaches approx $0.05\%$ of the total population.
\begin{figure}[h]
\centering
\includegraphics[width=8cm]{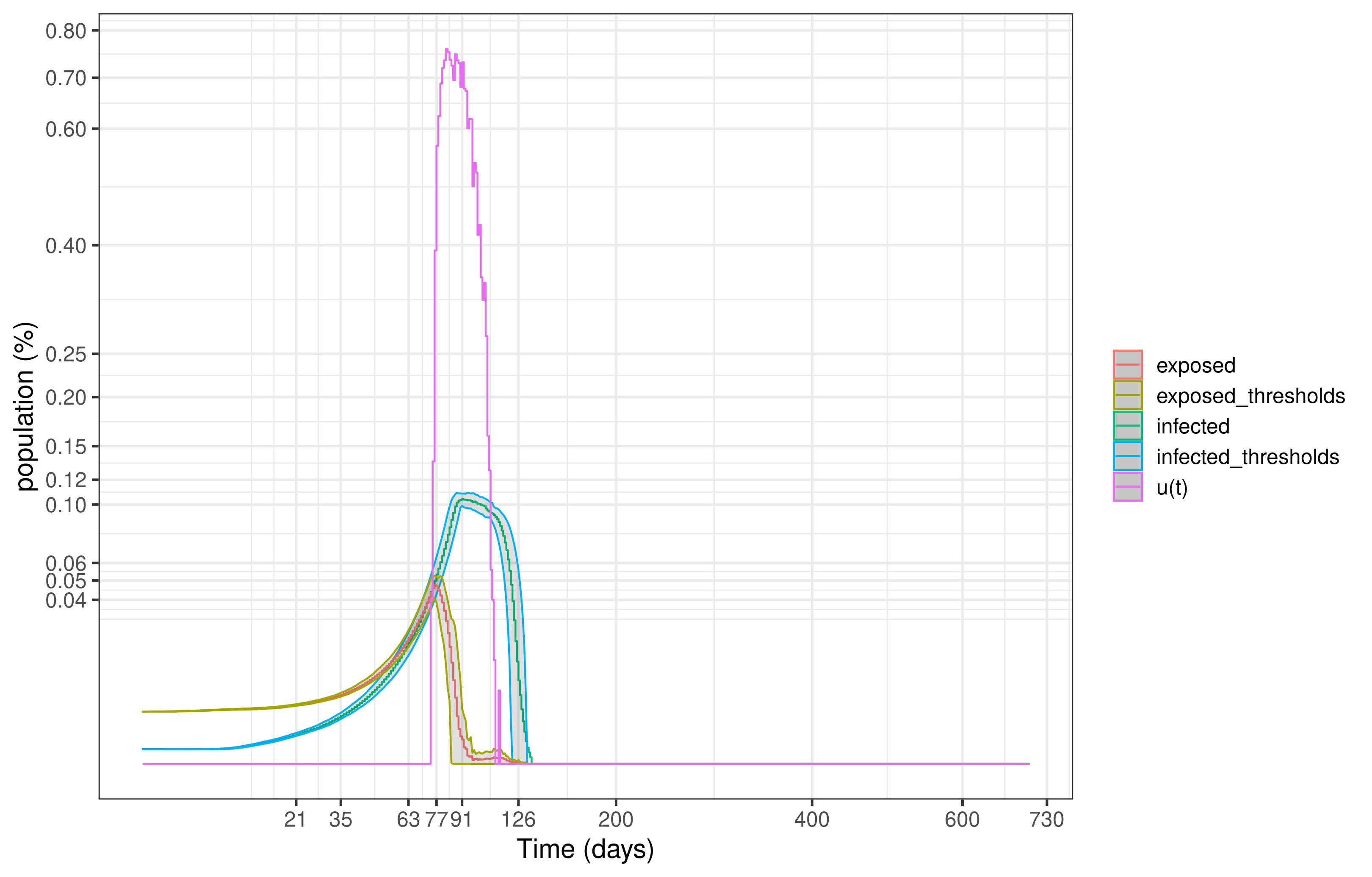}
\caption{CI E-I plot ($R_0\;=3.5$ control revised every day)}
\label{fig:CI E-I plot (R0=3.5 control revised every day)}
\end{figure}

Fig. \ref{fig:CI E-I plot (R0=3.5 control revised every day)} shows the CI plot for the exposed and infected populations. We have high control levels and a smooth behaviour as a result of changing the control every day.

\begin{figure}[H]
\centering
\includegraphics[width=8cm]{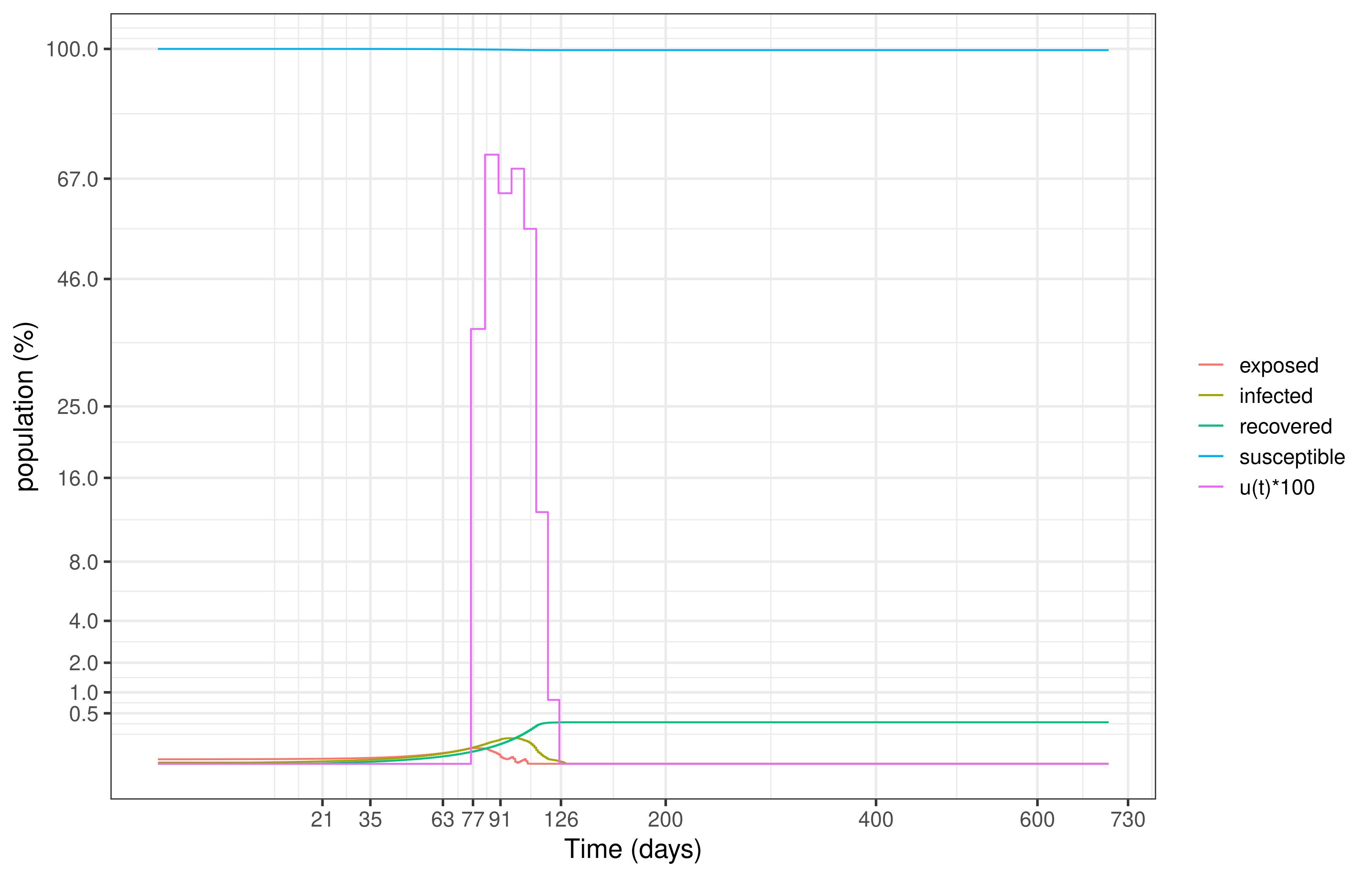}
\caption{SEIR plot ($R_0\;=3.5$ control revised every 7 days)}
\label{fig:SEIR plot(R0=3.5 control revised every 7 days}
\end{figure}
\begin{figure}[h]
\centering
\includegraphics[width=8cm]{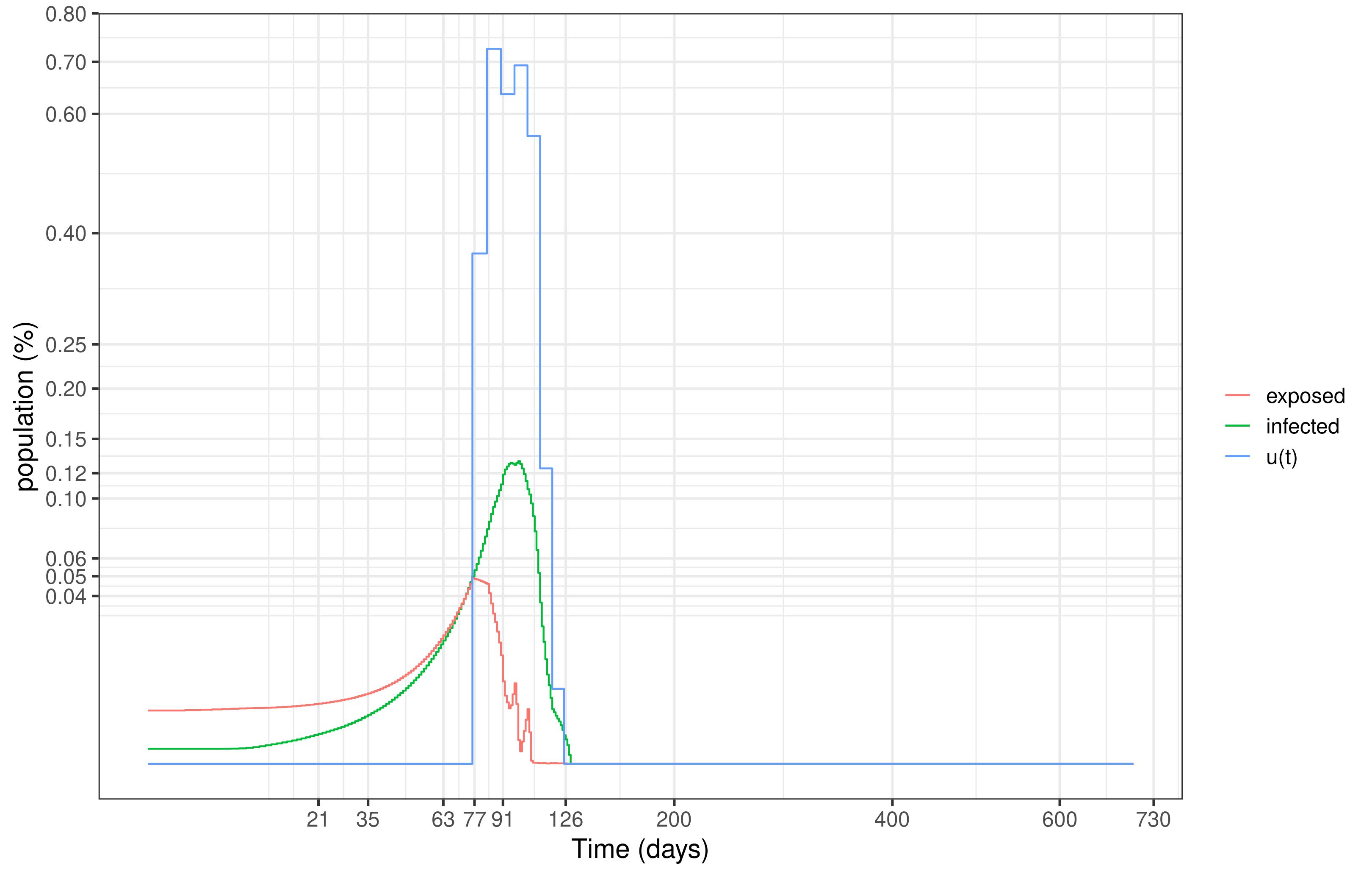}
\caption{Exposed-Infected plot ($R_0\;=3.5$ control revised every $7$ days}
\label{fig:Exposed-Infected plot(R0=3.5 control revised every 7 days}
\end{figure}

Fig. \ref{fig:SEIR plot(R0=3.5 control revised every 7 days} depicts the controlled trajectories for Case 2, with weekly control review and $R_0\;=3.5$. Fig. \ref{fig:Exposed-Infected plot(R0=3.5 control revised every 7 days} zooms in on the infected and exposed populations. We observe a slight increase in the peaks of infections and expositions as a result of the delayed control change. The 7-day review implies that control fluctuations in the form of overshoots and undershoots need to be corrected, but overall the outbreak is effectively mitigated. 
\begin{figure}[h]
\centering
\includegraphics[width=8cm]{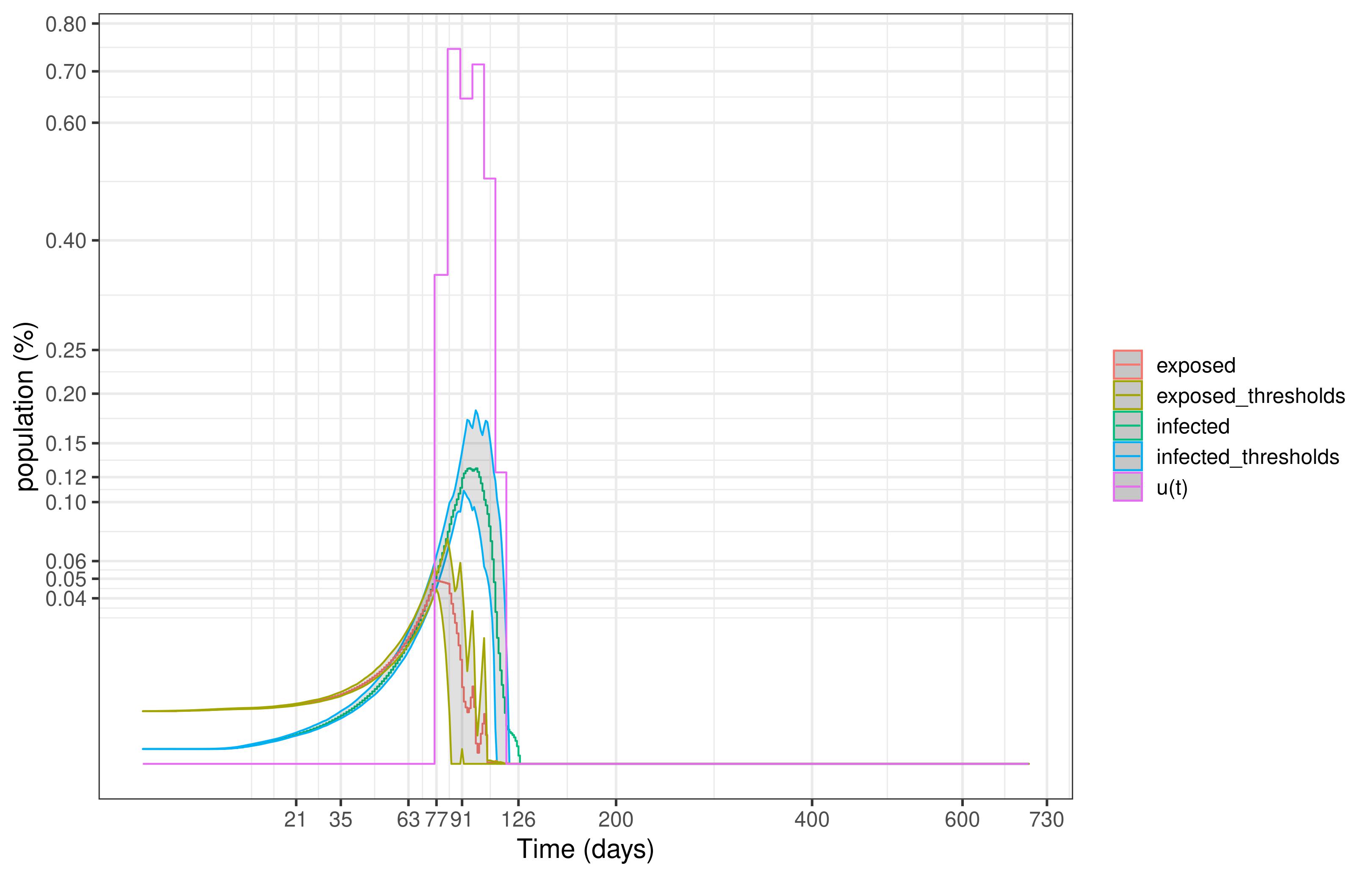}
\caption{CI E-I plot ($R_0\;=3.5$ control revised every $7$ days)}
\label{fig:CI E-I plot (R0=3.5 control revised every 7 day)}
\end{figure}

Fig. \ref{fig:CI E-I plot (R0=3.5 control revised every 7 day)} shows that the length of the CI increases with respect to Case 1 for both expositions and infections, as a result of the increased fluctuation due to the 7-day delay in control adjustment.

\begin{figure}[H]
\centering
\includegraphics[width=8cm]{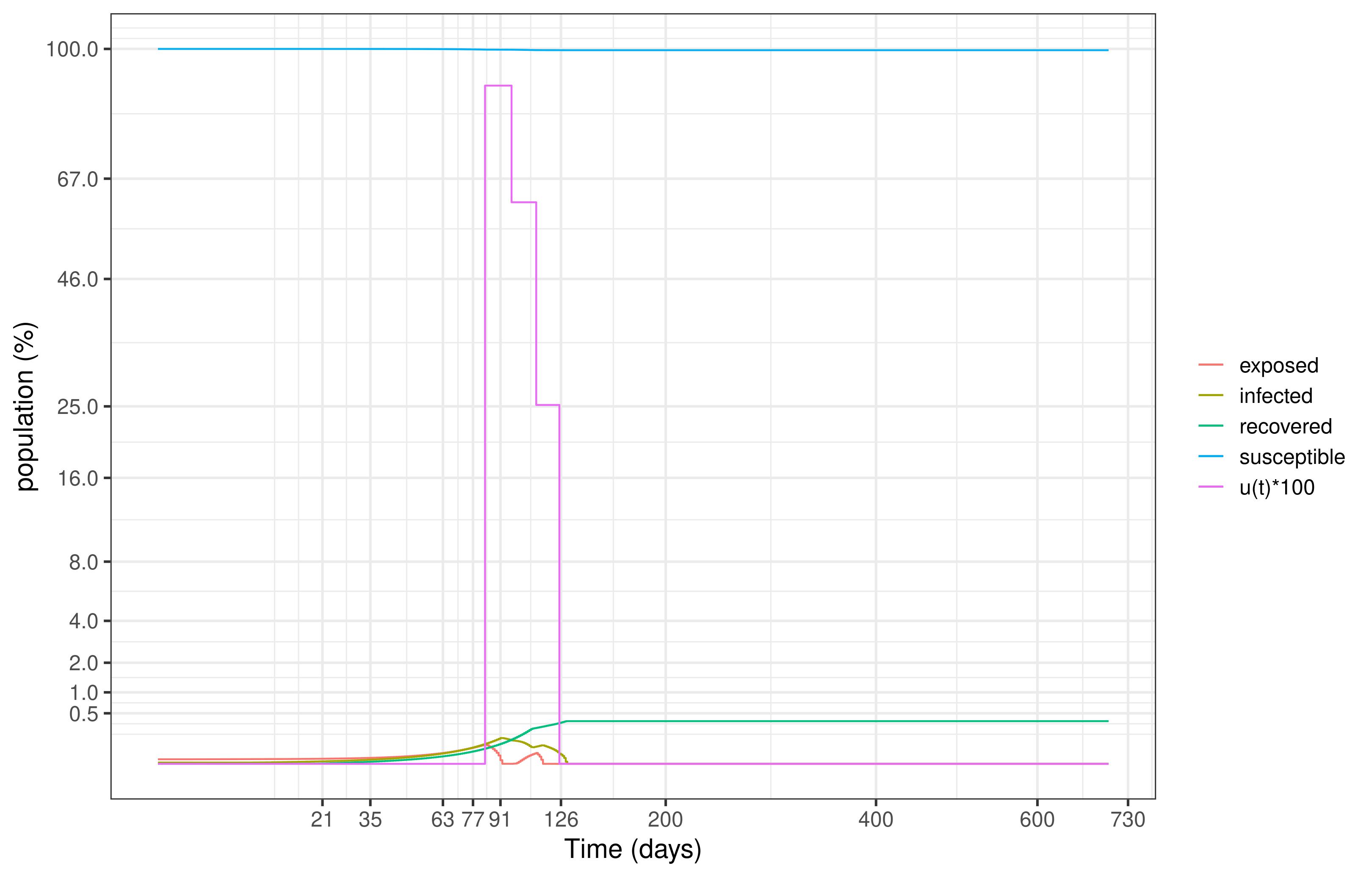}
\caption{SEIR plot ($R_0\;=3.5$ control revised every 14 days)}
\label{fig:SEIR plot(R0=3.5 control revised every 14 days}
\end{figure}

\begin{figure}[h]
\centering
\includegraphics[width=8cm]{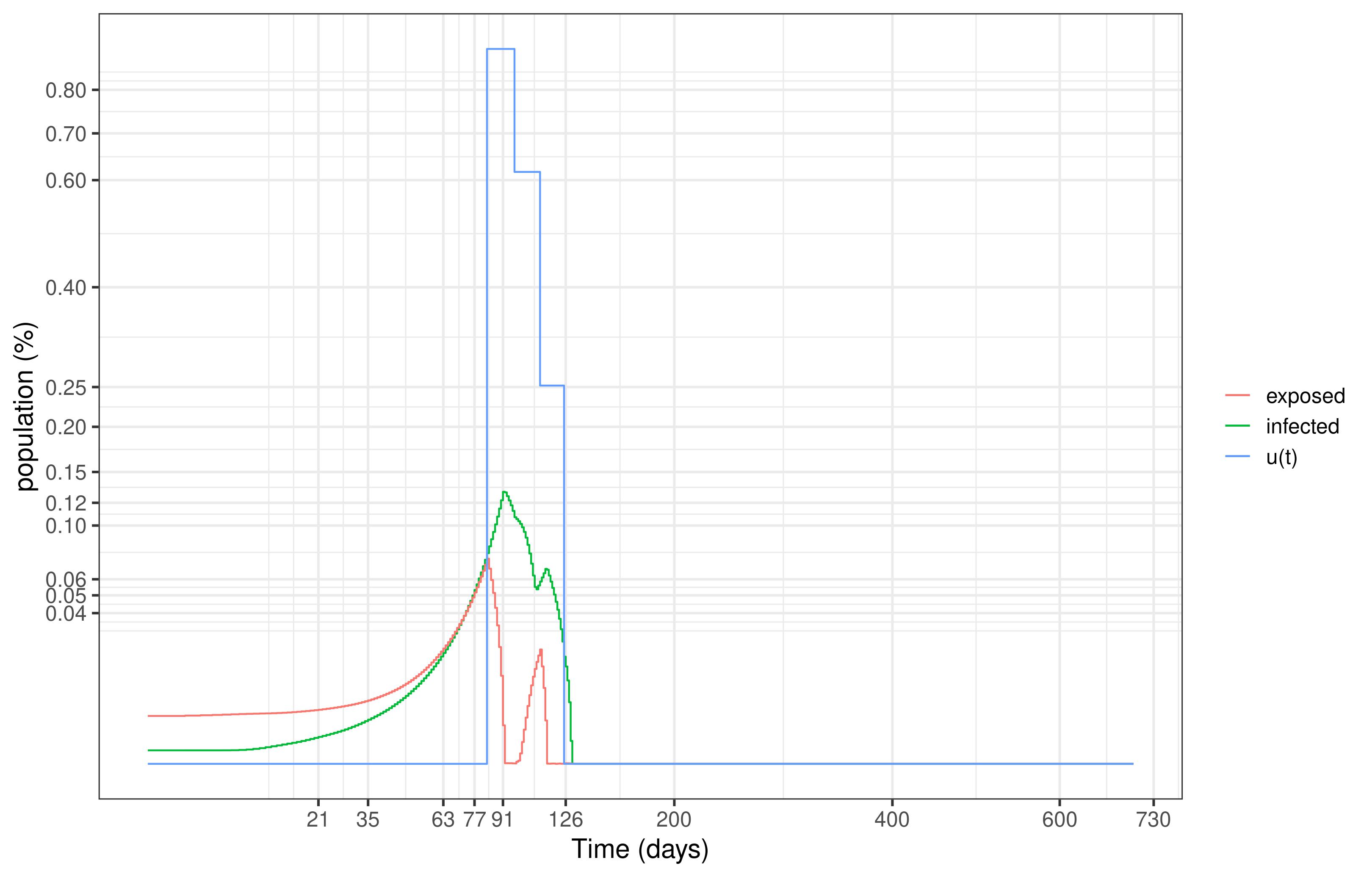}
\caption{Exposed-Infected plot ($R_0\;=3.5$ control revised every $14$ days)}
\label{fig:Exposed-Infected plot(R0=3.5 control revised every 14 days}
\end{figure}

Next we conduct the experiments for Case 3, with $R_0\;=3.5$ and control measures changing after every $14$ days. Fig. \ref{fig:SEIR plot(R0=3.5 control revised every 14 days} and  Fig. \ref{fig:Exposed-Infected plot(R0=3.5 control revised every 14 days} show a decrease in the peak of infections with respect to Case 2. Since high levels of control are enforced from the beginning, the number of infections remain limited until the policy is revised in the next period.
%
\begin{figure}[h]
\centering
\includegraphics[width=8cm]{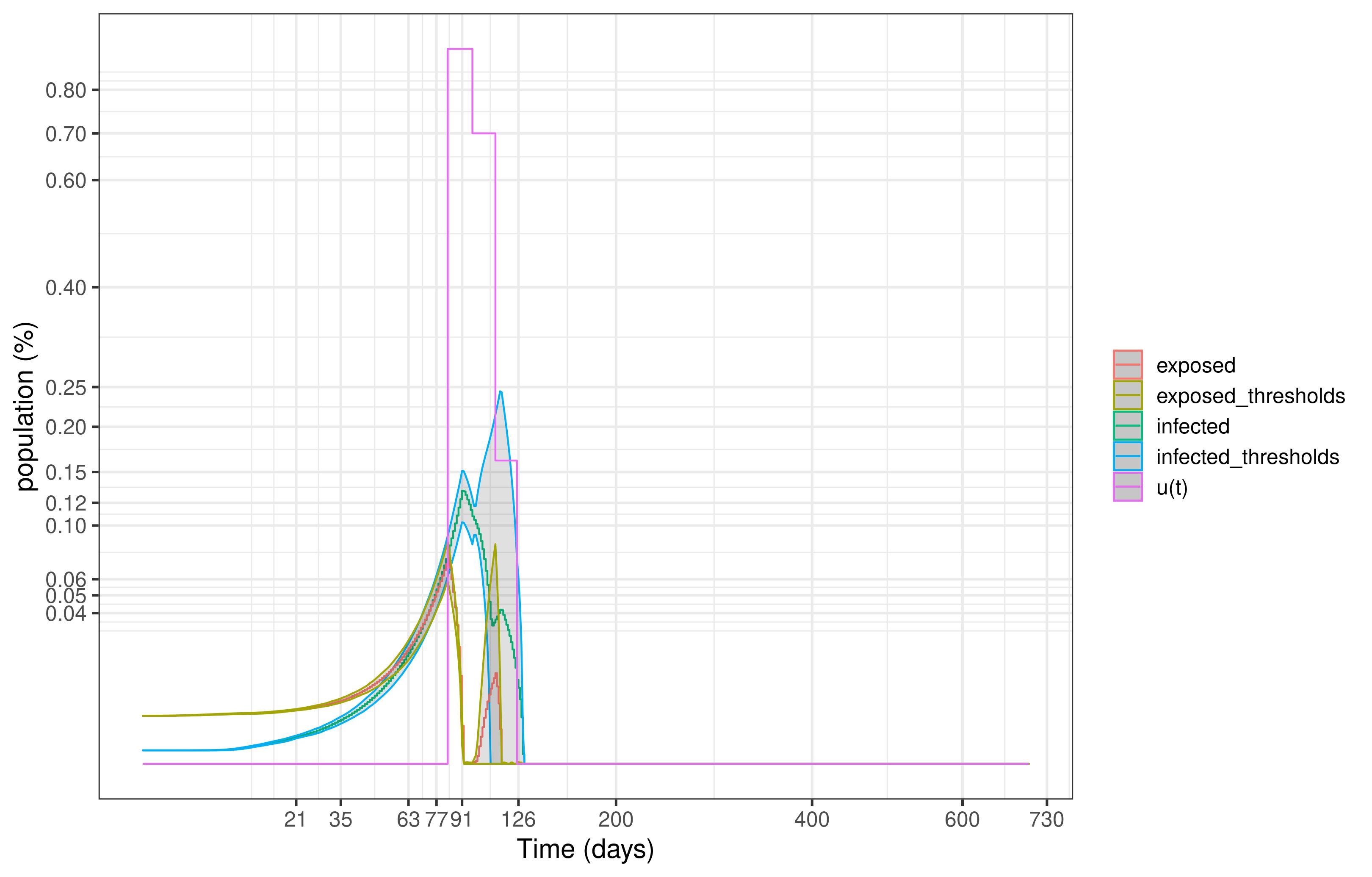}
\caption{CI E-I plot ($R_0\;=3.5$ control revised every $14$ days)}
\label{fig:CI E-I plot (R0=3.5 control revised every 14 day)}
\end{figure}

Fig. \ref{fig:CI E-I plot (R0=3.5 control revised every 14 day)} shows the CI plot and we observe that the 14-day review leads to a large variation in the infected and exposed populations across different realisations of the system. We see that the large review interval leads to increased volatility.



\begin{figure}[H]
\centering
\includegraphics[width=8cm]{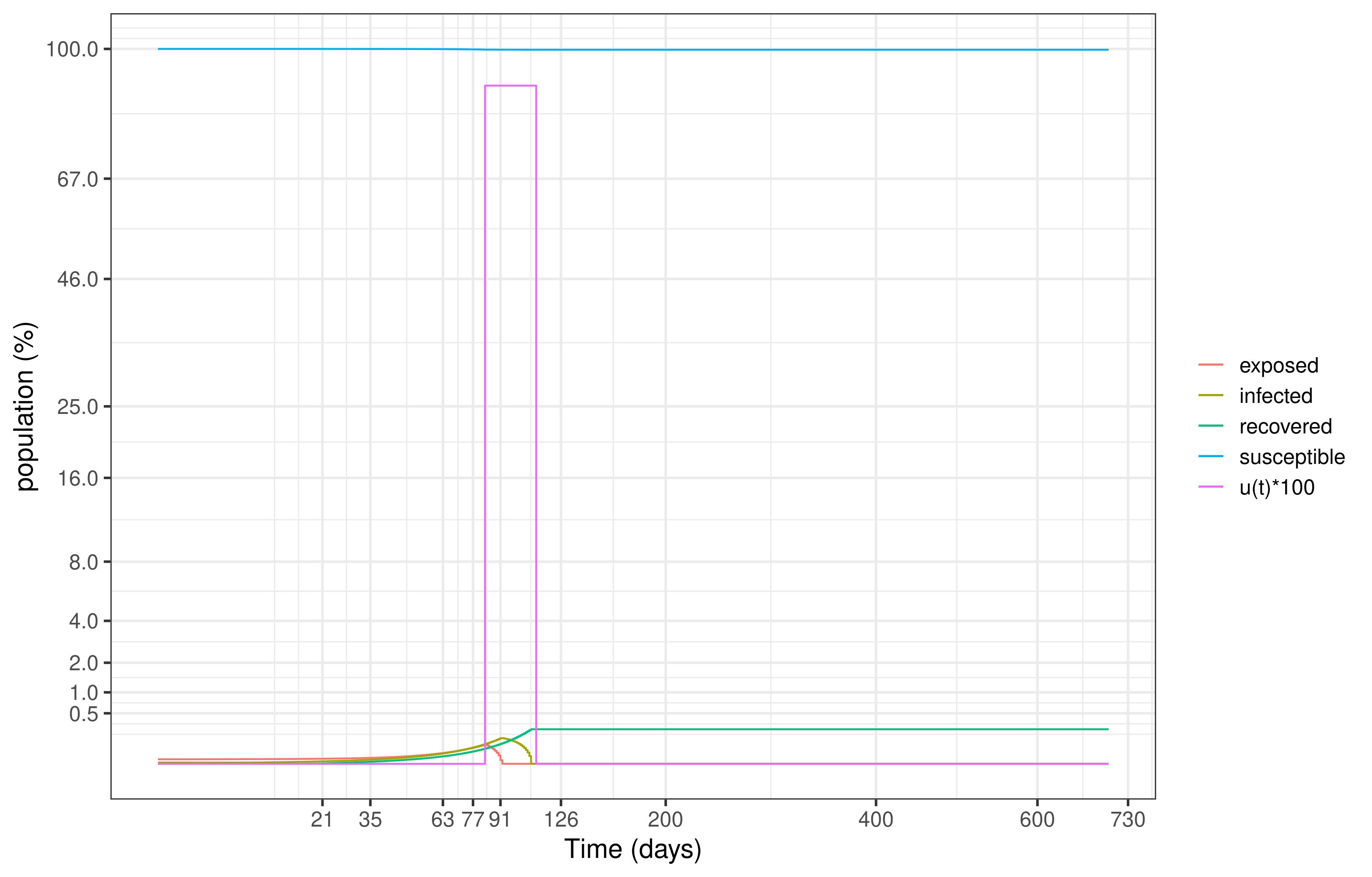}
\caption{SEIR plot ($R_0\;=3.5$ control revised every $28$ days)}
\label{fig:SEIR plot(R0=3.5 control revised every 28 days}
\end{figure}

\begin{figure}[h]
\centering
\includegraphics[width=8cm]{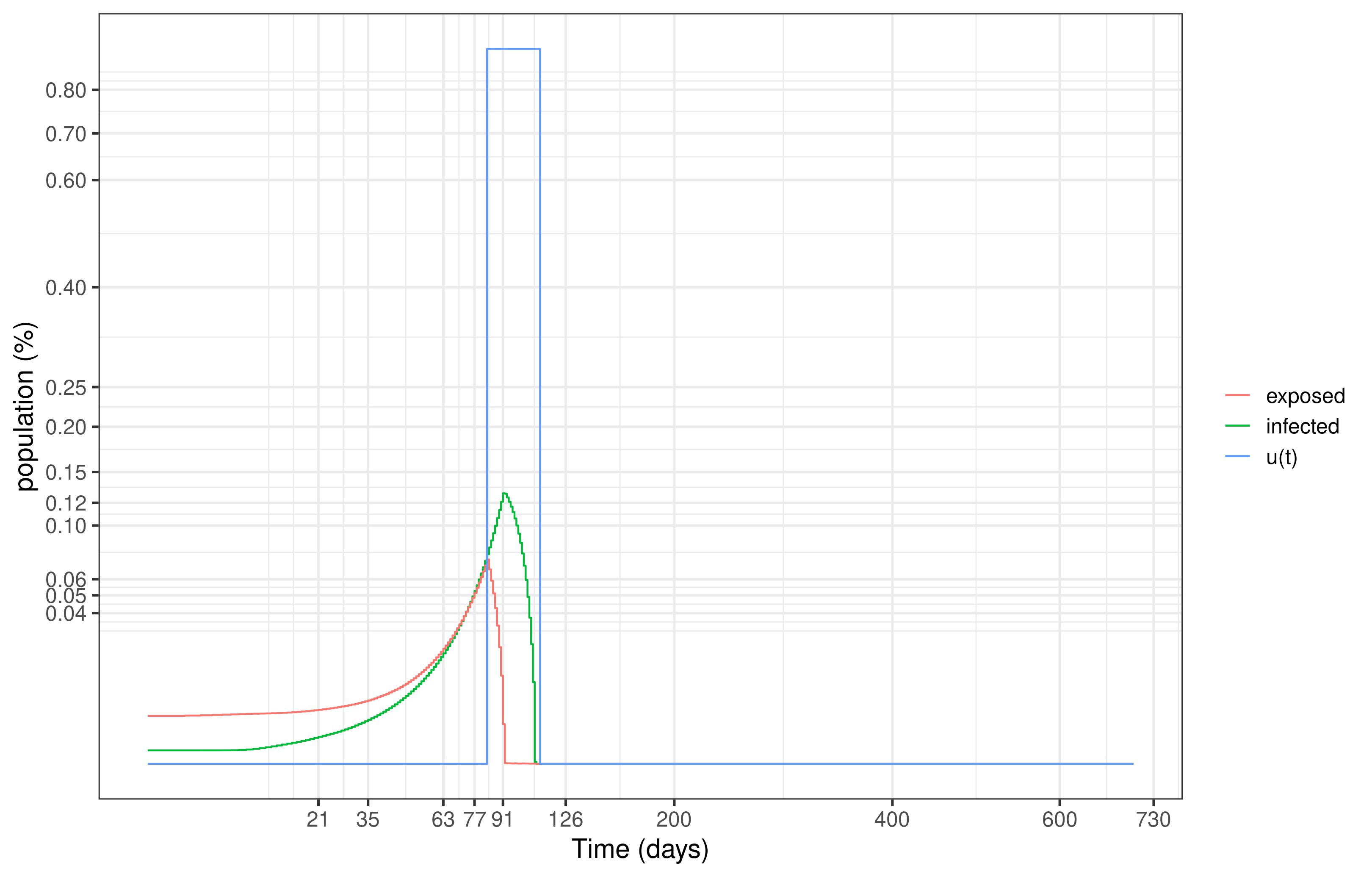}
\caption{Exposed-Infected plot ($R_0\;=3.5$ control revised every $28$ days)}
\label{fig:Exposed-Infected plot(R0=3.5 control revised every 28 days}
\end{figure}

The last set of experiments is for Case 4, with $R_0\;=3.5 $ and control measures changing after every $28$ days. Fig. \ref{fig:SEIR plot(R0=3.5 control revised every 28 days} and Fig. \ref{fig:Exposed-Infected plot(R0=3.5 control revised every 28 days} show the controlled trajectories. The control sequence curbs the epidemic and maintains reduced infection levels, similar to those in Cases 1 to 3.
\begin{figure}[h]
\centering
\includegraphics[width=8cm]{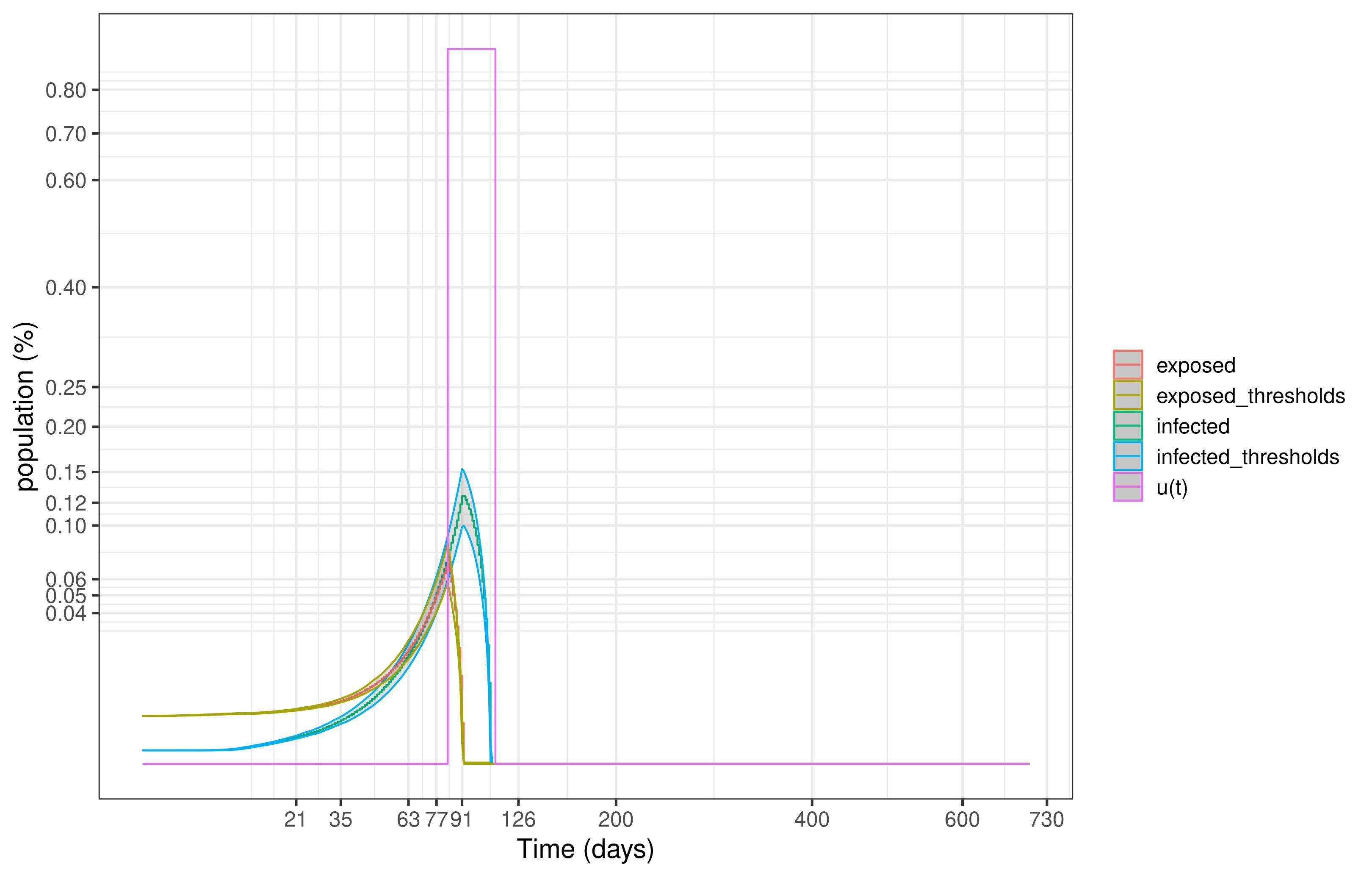}
\caption{Confidence-Interval E-I plot ($R_0\;=3.5$ control revised every $28$ days)}
\label{fig:CI E-I plot (R0=3.5 control revised every 28 day)}
\end{figure}

Fig. \ref{fig:CI E-I plot (R0=3.5 control revised every 28 day)} shows the CI for infections and expositions in Case 4. Since the control levels are high from the beginning, they are kept high for the first 28 days to stabilize the epidemic. The result is a decrease in the variance of the trajectories. It suggests that high control strategies are robust and can be used to ensure a decreasing epidemic when the control review period is large. Indeed, we can see that the trajectories with 14-day and 28-day review periods are rather similar and this is due to the high levels of control from the outset of the epidemic. This underscores the value of continuously monitoring an outbreak and revising the control as frequently as possible.


Finally, it is worth highlighting that, although large review periods do not compromise the ability to stabilise the epidemic, they force the decision maker to apply higher levels of control to conquer the outbreak. Indeed, by comparing the control levels for Cases 1 to 4, we can clearly notice an increase in the levels of control.

\balance

\section{Conclusions \label{sec:conc}}
This paper explored a parsimonious stochastic model to describe the evolution of an epidemic by means of the classical SEIR framework. Based on queues with infinite service capacity, the model is tractable and seamlessly incorporates general latency and infectious periods. It introduces a stochastic optimal control formulation that seeks an optimal trade-off between the occupation of the healthcare system and the economic impacts of mitigation measures. Since the model is based  on  parsimonious  stochastic  formulations,  it  is tractable, easy to use and guaranteed to converge. 

We tested the formulation on the epidemic data from New Delhi, including experiments with $R_0=2.5$ and $R_0=3.5$ to account for data uncertainty. The results show that an optimal control policy acts swiftly in the first wave of the epidemic, therefore avoiding successive waves as observed in countries where control was delayed or relaxed too soon. They also highlight that continuous monitoring is vital, as larger policy review periods imply higher levels of control to curb the epidemic. It is apparent that failing to act swiftly in the first wave is more costly, since myopically sacrificing the healthcare system to preserve the economy in the outset of the epidemic will require higher levels of control to contain a larger outbreak. Furthermore, failing to act swiftly may lead to the collapse of the healthcare system, thereby incurring additional economic and societal costs.

\bibliographystyle{plainnat}
\bibliography{references}

\begin{thebibliography}{43}
\providecommand{\natexlab}[1]{#1}
\providecommand{\url}[1]{\texttt{#1}}
\expandafter\ifx\csname urlstyle\endcsname\relax
  \providecommand{\doi}[1]{doi: #1}\else
  \providecommand{\doi}{doi: \begingroup \urlstyle{rm}\Url}\fi

\bibitem[Allen(2008)]{Allen2008}
L.~J.~S. Allen.
\newblock {An Introduction to Stochastic Epidemic Models}.
\newblock In \emph{Mathematical Epidemiology}, pages 81--130. Springer Berlin
  Heidelberg, 2008.

\bibitem[Amador and Lopez-Herrero(2018)]{Amador2018}
J.~Amador and M.~Lopez-Herrero.
\newblock {Cumulative and Maximum Epidemic Sizes for A Nonlinear SEIR
  Stochastic Model with Limited Resources}.
\newblock \emph{Discrete \& Continuous Dynamical Systems - B}, 23\penalty0
  (8):\penalty0 3137--3151, 2018.

\bibitem[Angelo et~al.(2017)Angelo, Arruda, Goldwasser, Lobo, Salles, and
  Silva]{Angelo2017}
S.~A. Angelo, E.~F. Arruda, R.~S. Goldwasser, M.~S.~C. Lobo, A.~A. Salles, and
  J.~R.~L. Silva.
\newblock {Demand Forecast and Optimal Planning Of Intensive Care Unit (ICU)
  Capacity}.
\newblock \emph{Pesquisa Operacional}, 37\penalty0 (2):\penalty0 229 -- 245, 08
  2017.

\bibitem[Arruda et~al.(2021{\natexlab{a}})Arruda, Das, Dias, and
  Pastore]{Arruda2021-Plos}
E.~F. Arruda, S.~S. Das, C.~M. Dias, and D.~H. Pastore.
\newblock {Modelling and Optimal Control of Multi Strain Epidemics, with
  Application to COVID-19}.
\newblock \emph{PLOS ONE}, 16\penalty0 (9):\penalty0 1--18, 09
  2021{\natexlab{a}}.

\bibitem[Arruda et~al.(2021{\natexlab{b}})Arruda, e~A.~Alexandre, Fragoso,
  do~Val, and Thomas]{Arruda2021novel}
E.~F. Arruda, R.~e~A.~Alexandre, M.~D. Fragoso, J.~B.~R. do~Val, and S.~S.
  Thomas.
\newblock {A Novel Stochastic Epidemic Model with Application to COVID-19}.
\newblock \emph{CoRR}, abs/2102.08213, 2021{\natexlab{b}}.

\bibitem[Artalejo et~al.(2015)Artalejo, Economou, and
  Lopez-Herrero]{Artalejo2015}
J.~R. Artalejo, A.~Economou, and M.~J. Lopez-Herrero.
\newblock {The Stochastic SEIR Model Before Extinction: Computational
  Approaches}.
\newblock \emph{Applied Mathematics and Computation}, 265\penalty0
  (C):\penalty0 1026 -- 1043, 2015.

\bibitem[Backer et~al.(2020)Backer, Klinkenberg, and Wallinga]{Backer2020}
J.~A. Backer, D.~Klinkenberg, and J.~Wallinga.
\newblock {Incubation Period of 2019 Novel Coronavirus {(2019-nCoV)} Infections
  among Travellers from Wuhan, China, 20–28 January 2020}.
\newblock \emph{Eurosurveillance}, 25\penalty0 (5), Feb 2020.

\bibitem[Bergonzi et~al.(2021)Bergonzi, Pecker-Marcosig, Kofman, and
  Castro]{bergonzi2020discrete}
Mariana Bergonzi, Ezequiel Pecker-Marcosig, Ernesto Kofman, and Rodrigo~Daniel
  Castro.
\newblock {Discrete-Time Modeling of COVID-19 Propagation in Argentina with
  Explicit Delays}.
\newblock \emph{Comput. Sci. Eng.}, 23\penalty0 (1):\penalty0 35--45, 2021.

\bibitem[Bertozzi et~al.(2020)Bertozzi, Franco, Mohler, Short, and
  Sledge]{Bertozzi2020}
Andrea~L. Bertozzi, Elisa Franco, George Mohler, Martin~B. Short, and Daniel
  Sledge.
\newblock {The Challenges of Modeling and Forecasting the Spread of COVID-19}.
\newblock \emph{Proc. of the National Academy of Sciences}, 117\penalty0
  (29):\penalty0 16732--16738, 2020.

\bibitem[Bootsma and Ferguson(2007)]{Bootsma2007}
Martin C.~J. Bootsma and Neil~M. Ferguson.
\newblock {The Effect of Public Health Measures on The 1918 Influenza Pandemic
  in U.S. Cities}.
\newblock \emph{Proc. of the National Academy of Sciences}, 104\penalty0
  (18):\penalty0 7588--7593, 2007.

\bibitem[Britton(2010)]{Britton2010}
T.~Britton.
\newblock {Stochastic Epidemic Models: A Survey}.
\newblock \emph{Mathematical Biosciences}, 225\penalty0 (1):\penalty0 24 -- 35,
  May 2010.

\bibitem[Cheetham et~al.(2021)Cheetham, Waites, Ebyarimpa, Leber, Brennan, and
  Panovska-Griffiths]{cheetham2020determining}
Nathan Cheetham, William Waites, Irene Ebyarimpa, Werner Leber, Katie Brennan,
  and Jasmina Panovska-Griffiths.
\newblock {Determining The Level of Social Distancing Necessary to Avoid A
  Second COVID-19 Epidemic Wave: A Modelling Study for North East London}.
\newblock \emph{Scientific Reports}, 11\penalty0 (1):\penalty0 5806, 2021.

\bibitem[Clancy(2014)]{Clancy2014}
D.~Clancy.
\newblock {SIR Epidemic Models with General Infectious Period Distribution}.
\newblock \emph{Statistics \& Probability Letters}, 85:\penalty0 1 -- 5, Feb
  2014.

\bibitem[Dyer(2021)]{Dyern2021}
Owen Dyer.
\newblock {COVID-19: Peru's Official Death Toll Triples to Become World's
  Highest}.
\newblock \emph{BMJ}, 373, 2021.

\bibitem[Eick et~al.(1993)Eick, Massey, and Whitt]{Eick1993}
S.~G. Eick, W.~A. Massey, and W.~Whitt.
\newblock {The Physics of the $M_t/G/\infty$ Queue}.
\newblock \emph{Operations Research}, 41\penalty0 (4):\penalty0 731--742, 1993.

\bibitem[Ferguson et~al.(2020)Ferguson, Laydon, Nedjati-Gilani, Imai, Ainslie,
  Baguelin, Bhatia, Boonyasiri, Cucunubá, Cuomo-Dannenburg, Dighe, Dorigatti,
  Fu, Gaythorpe, Green, Hamlet, Hinsley, Okell, van Elsland, Thompson, Verity,
  Volz, Wang, Wang, Walker, Walters, Winskill, Whittaker, Donnelly, Riley, and
  Ghani]{ferguson2020}
Neil~M Ferguson, Daniel Laydon, Gemma Nedjati-Gilani, Natsuko Imai, Kylie
  Ainslie, Marc Baguelin, Sangeeta Bhatia, Adhiratha Boonyasiri, Zulma
  Cucunubá, Gina Cuomo-Dannenburg, Amy Dighe, Ilaria Dorigatti, Han Fu, Katy
  Gaythorpe, Will Green, Arran Hamlet, Wes Hinsley, Lucy~C Okell, Sabine van
  Elsland, Hayley Thompson, Robert Verity, Erik Volz, Haowei Wang, Yuanrong
  Wang, Patrick~GT Walker, Caroline Walters, Peter Winskill, Charles Whittaker,
  Christl~A Donnelly, Steven Riley, and Azra~C Ghani.
\newblock {Report 9: Impact of Non-pharmaceutical Interventions (NPIs) to
  Reduce {COVID-19} Mortality and Healthcare Demand}.
\newblock Technical report, Imperial College London, 03 2020.

\bibitem[Goldwasser et~al.(2016)Goldwasser, Lobo, Arruda, Angelo, Silva,
  Salles, and David]{Goldwasser2016}
R.~S. Goldwasser, M.~S.~C. Lobo, E.~F. Arruda, S.~A. Angelo, J.~R.~L. Silva,
  A.~A. Salles, and C.~M. David.
\newblock {Difficulties in Access and Estimates of Public Beds in Intensive
  Care Units in the State of Rio de Janeiro}.
\newblock \emph{Revista de Saude Publica}, 50, 2016.
\newblock ISSN 0034-8910.

\bibitem[Gross et~al.(2018)Gross, Shortle, Thompson, and Harris]{Shortle2018}
Donald Gross, John~F. Shortle, James~M. Thompson, and Carl~M. Harris.
\newblock \emph{{Fundamentals of Queuing Theory}}.
\newblock Wiley Series in Probability and Statistics. 5 edition, 2018.

\bibitem[Gómez-Corral and López-García(2017)]{Corral2017}
A.~Gómez-Corral and M.~López-García.
\newblock {On SIR Epidemic Models with Generally Distributed Infectious
  Periods: Number of Secondary Cases and Probability of Infection}.
\newblock \emph{International Journal of Biomathematics}, 10\penalty0
  (2):\penalty0 1750024, 2017.

\bibitem[Kantner and Koprucki(2020)]{Kantner2020}
M.~Kantner and T.~Koprucki.
\newblock {Beyond Just ``Flattening The Curve'': Optimal Control of Epidemics
  with Purely Non-pharmaceutical Interventions}.
\newblock \emph{Journal of Mathematics in Industry}, 10:\penalty0 23, 2020.

\bibitem[Kermack and McKendrick(1927)]{Kermack1927}
W.~O. Kermack and A.~G. McKendrick.
\newblock {A Contribution to the Mathematical Theory of Epidemics}.
\newblock \emph{Proc. of the Royal Society of London. Series A, Containing
  Papers of a Mathematical and Physical Character}, 115\penalty0
  (772):\penalty0 700--721, 1927.

\bibitem[Latif et~al.(2020)Latif, Usman, Manzoor, Iqbal, Qadir, Tyson, Castro,
  Razi, Boulos, Weller, and Crowcroft]{Latif2020}
Siddique Latif, Muhammad Usman, Sanaullah Manzoor, Waleed Iqbal, Junaid Qadir,
  Gareth Tyson, Ignacio Castro, Adeel Razi, Maged N.~Kamel Boulos, Adrian
  Weller, and Jon Crowcroft.
\newblock {Leveraging Data Science to Combat COVID-19: A Comprehensive Review}.
\newblock \emph{IEEE Trans. on Artificial Intelligence}, 1\penalty0
  (1):\penalty0 85--103, 2020.

\bibitem[Lemos et~al.(2020)Lemos, D'Angelo, Farias, Almeida, Gomes, Pinto,
  {Cavalcante Filho}, Feijao, Cardoso, Lima, Linhares, Mello, Coelho, and
  Cavalcanti]{Lemos2020}
D.~R.~Q. Lemos, S.~M. D'Angelo, L.A.B.G. Farias, M.~M. Almeida, R.~G. Gomes,
  G.~P. Pinto, J.~N. {Cavalcante Filho}, L.~X. Feijao, A.~R.~P. Cardoso,
  T.~B.~R. Lima, P.~M.~C. Linhares, L.~P. Mello, T.~M. Coelho, and L.~P.~G.
  Cavalcanti.
\newblock {Health System Collapse 45 days after the Detection of COVID-19 in
  Ceara, Northeast Brazil: A Preliminary Analysis}.
\newblock \emph{Revista da Sociedade Brasileira de Medicina Tropical}, 53,
  2020.

\bibitem[Lopez-Herrero(2016)]{Lopez2017}
Mariajesus Lopez-Herrero.
\newblock {Epidemic Transmission on SEIR Stochastic Models with Nonlinear
  Incidence Rate}.
\newblock \emph{Mathematical Methods in the Applied Sciences}, 40\penalty0
  (7):\penalty0 2532--2541, 2016.

\bibitem[Mandal et~al.(2020)Mandal, Jana, Nandi, Khatua, Adak, and
  Kar]{RePEc:eee:chsofr:v:136:y:2020:i:c:s0960077920302897}
Manotosh Mandal, Soovoojeet Jana, Swapan~Kumar Nandi, Anupam Khatua, Sayani
  Adak, and T.~K. Kar.
\newblock {A Model based Study on The Dynamics of COVID-19: Prediction and
  Control}.
\newblock \emph{Chaos, Solitons \& Fractals}, 136\penalty0 (C), 2020.

\bibitem[Marimuthu et~al.(2021)Marimuthu, Joy, Malavika, Nadaraj, Asirvatham,
  and Jeyaseelan]{S.MariMuthu2020}
S.~Marimuthu, Melvin Joy, B.~Malavika, Ambily Nadaraj, Edwin~Sam Asirvatham,
  and L.~Jeyaseelan.
\newblock {Modelling of Reproduction Number for COVID-19 in India and High
  Incidence States}.
\newblock \emph{Clinical Epidemiology and Global Health}, 9:\penalty0 57--61,
  2021.

\bibitem[Masum et~al.(2020)Masum, Shahriar, Haddad, and Alam]{masum2020r}
Mohammad Masum, Hossain Shahriar, Hisham~M Haddad, and Md~Shafiul Alam.
\newblock {r-LSTM: Time Series Forecasting for COVID-19 Confirmed Cases with
  LSTM based Framework}.
\newblock In \emph{IEEE International Conference on Big Data}, pages
  1374--1379, 2020.

\bibitem[Meidan et~al.(2021)Meidan, Schulmann, Cohen, Haber, Yaniv, Sarid, and
  Barzel]{meidan2021alternating}
Dror Meidan, Nava Schulmann, Reuven Cohen, Simcha Haber, Eyal Yaniv, Ronit
  Sarid, and Baruch Barzel.
\newblock {Alternating Quarantine for Sustainable Epidemic Mitigation}.
\newblock \emph{Nature Communications}, 12\penalty0 (1):\penalty0 1--12, Jan
  2021.

\bibitem[Meyerowitz et~al.(2021)Meyerowitz, Richterman, Bogoch, Low, and
  Cevik]{Meyerowitz2020}
Eric~A Meyerowitz, Aaron Richterman, Isaac~I Bogoch, Nicola Low, and Muge
  Cevik.
\newblock {Towards An Accurate and Systematic Characterisation of Persistently
  Asymptomatic Infection with SARS-CoV-2}.
\newblock \emph{The Lancet Infectious Diseases}, 21\penalty0 (6):\penalty0
  163--169, Jun 2021.

\bibitem[OECD(2018)]{OECD2018}
OECD.
\newblock {Hospital Beds}.
\newblock 2018.
\newblock URL \url{https://www.oecd-ilibrary.org/content/data/0191328e-en}.

\bibitem[Ohi et~al.(2020)Ohi, Mridha, Monowar, and Hamid]{ohi2020exploring}
Abu~Quwsar Ohi, MF~Mridha, Muhammad~Mostafa Monowar, and Md~Abdul Hamid.
\newblock {Exploring Optimal Control of Epidemic Spread Using Reinforcement
  Learning}.
\newblock \emph{Scientific Reports}, 10\penalty0 (1):\penalty0 1--19, 2020.

\bibitem[Oraby et~al.(2021)Oraby, Tyshenko, Maldonado, Vatcheva, Elsaadany,
  Alali, Longenecker, and Al-Zoughool]{Oraby2021}
Tamer Oraby, Michael~G. Tyshenko, Jose~Campo Maldonado, Kristina Vatcheva,
  Susie Elsaadany, Walid~Q. Alali, Joseph~C. Longenecker, and Mustafa
  Al-Zoughool.
\newblock {Modeling The Effect of Lockdown Timing as a COVID-19 Control Measure
  in Countries with Differing Social Contacts}.
\newblock \emph{Scientific Reports}, 11\penalty0 (1):\penalty0 3354, Feb 2021.

\bibitem[Perkins and Espa{\~n}a(2020)]{Perkins2020}
T.~A. Perkins and G.~Espa{\~n}a.
\newblock {Optimal Control of the COVID-19 Pandemic with Non-pharmaceutical
  Interventions}.
\newblock \emph{Bulletin of Mathematical Biology}, 82\penalty0 (9):\penalty0
  118, Sep 2020.

\bibitem[Plazas et~al.(2021)Plazas, Malvestio, Starnini, and
  D{\'{\i}}az{-}Guilera]{plazas2021modeling}
Adri{\`{a}} Plazas, Irene Malvestio, Michele Starnini, and Albert
  D{\'{\i}}az{-}Guilera.
\newblock {Modeling Partial Lockdowns in Multiplex Networks using Partition
  Strategies}.
\newblock \emph{Applied Network Science}, 6\penalty0 (1):\penalty0 1--15, Mar
  2021.

\bibitem[Priesemann et~al.(2021{\natexlab{a}})Priesemann, Balling, Brinkmann,
  Ciesek, Czypionka, Eckerle, Giordano, Hanson, Hel, Hotulainen, Klimek,
  Nassehi, Peichl, Perc, Petelos, Prainsack, and Szczurek]{Priesemann2021}
Viola Priesemann, Rudi Balling, Melanie~M Brinkmann, Sandra Ciesek, Thomas
  Czypionka, Isabella Eckerle, Giulia Giordano, Claudia Hanson, Zdenek Hel,
  Pirta Hotulainen, Peter Klimek, Armin Nassehi, Andreas Peichl, Matjaz Perc,
  Elena Petelos, Barbara Prainsack, and Ewa Szczurek.
\newblock {An Action Plan for pan-European Defence Against New SARS-CoV-2
  Variants}.
\newblock \emph{The Lancet}, 397\penalty0 (10273):\penalty0 469--470,
  2021{\natexlab{a}}.

\bibitem[Priesemann et~al.(2021{\natexlab{b}})Priesemann, Brinkmann, Ciesek,
  Cuschieri, Czypionka, Giordano, Gurdasani, Hanson, Hens, Iftekhar,
  Kelly-Irving, Klimek, Kretzschmar, Peichl, Perc, Sannino, Schernhammer,
  Schmidt, Staines, and Szczurek]{Priesemann2021a}
Viola Priesemann, Melanie Brinkmann, Sandra Ciesek, Sarah Cuschieri, Thomas
  Czypionka, Giulia Giordano, Deepti Gurdasani, Claudia Hanson, Niel Hens, Emil
  Iftekhar, Michelle Kelly-Irving, Peter Klimek, Mirjam Kretzschmar, Andreas
  Peichl, Matjaž Perc, Francesco Sannino, Eva Schernhammer, Alexander Schmidt,
  Anthony Staines, and Ewa Szczurek.
\newblock {Calling for Pan-European Commitment for Rapid and Sustained
  Reduction in SARS-CoV-2 Infections}.
\newblock \emph{The Lancet}, 397\penalty0 (10269):\penalty0 92--93,
  2021{\natexlab{b}}.

\bibitem[Puterman(1994)]{Puterman1994}
M.L. Puterman.
\newblock \emph{{Markov Decision Processes: Discrete Stochastic Dynamic
  Programming}}.
\newblock John Wiley {\&} Sons, Inc., 1st edition, 1994.

\bibitem[Ramachandran(2020)]{Ramachandran2020}
R.~Ramachandran.
\newblock {COVID-19 — A Very Visible Pandemic}.
\newblock \emph{The Lancet}, 396\penalty0 (10248):\penalty0 e13--e14, Aug 2020.

\bibitem[Ross(1916)]{Ross1916}
R.~Ross.
\newblock {An Application of the Theory of Probabilities to the Study of a
  Priori Pathometry-Part I}.
\newblock \emph{Proc. of the Royal Society of London. Series A, Containing
  Papers of a Mathematical and Physical Character}, 92\penalty0 (638):\penalty0
  204--230, 1916.

\bibitem[Sabino et~al.(2021)Sabino, Buss, Carvalho, {Prete Jr}t, Crispim,
  Fraiji, Pereira, Parag, Peixoto, Kraemer, Oikawa, Salomon, Cucunuba, Castro,
  Santos, Nascimento, Pereira, Ferguson, Pybus, Kucharski, Busch, Dye, and
  Faria]{Sabino2021}
E.~C. Sabino, L.~F. Buss, M.~P.~S. Carvalho, C.~A. {Prete Jr}t, M.~A.~E.
  Crispim, N.~A. Fraiji, R.~H.~M. Pereira, K.~V. Parag, P.~S. Peixoto, M.~U.~G.
  Kraemer, M.~K. Oikawa, T.~Salomon, Z.~M. Cucunuba, M.~C. Castro, A.~A.~S.
  Santos, V.~H. Nascimento, H.~S. Pereira, N.~M. Ferguson, O.~G. Pybus,
  A.~Kucharski, M.~P. Busch, C.~Dye, and N.~R. Faria.
\newblock {Resurgence of COVID-19 in Manaus, Brazil, Despite High
  Seroprevalence}.
\newblock \emph{The Lancet}, 397\penalty0 (10273):\penalty0 452--455, 2021.

\bibitem[Tarrataca et~al.(2021)Tarrataca, Dias, Haddad, and
  Arruda]{Tarrataca2021}
L.~Tarrataca, C.~M. Dias, D.~Haddad, and E.~F. Arruda.
\newblock {Flattening the Curves: On-off Lock-down Strategies for COVID-19 with
  an Application to Brazil}.
\newblock \emph{Journal of Mathematics in Industry}, 11\penalty0 (1), 2021.

\bibitem[Wallinga and Lipsitch(2007)]{Walling2007}
J.~Wallinga and M.~Lipsitch.
\newblock {How Generation Intervals Shape the Relationship between Growth Rates
  and Reproductive Numbers}.
\newblock \emph{Proc. of the Royal Society B: Biological Sciences},
  274\penalty0 (1609):\penalty0 599--604, 2007.

\bibitem[You et~al.(2020)You, Wang, Zhang, Song, Xu, and Lai]{You2020}
Shibing You, Hengli Wang, Miao Zhang, Haitao Song, Xiaoting Xu, and Yongzeng
  Lai.
\newblock {Assessment of Monthly Economic Losses in Wuhan under The Lockdown
  Against COVID-19}.
\newblock \emph{Humanities and Social Sciences Communications}, 7\penalty0
  (1):\penalty0 1--12, 2020.

\end{thebibliography}


\end{document}